\documentclass{article}
\title{Permutational Quantum Computing}
\author{Stephen P. Jordan \\
\small {\em Institute for Quantum Information, California Institute
 of Technology.} \texttt{sjordan@caltech.edu} \small}

\date{}

\usepackage{graphicx, amssymb, amsmath, bm, fullpage, dsfont}

\input{Qcircuit}

\newcommand{\captionfonts}{\small}

\makeatletter  
\long\def\@makecaption#1#2{%
  \vskip\abovecaptionskip
  \sbox\@tempboxa{{\captionfonts #1: #2}}%
  \ifdim \wd\@tempboxa >\hsize
    {\captionfonts #1: #2\par}
  \else
    \hbox to\hsize{\hfil\box\@tempboxa\hfil}%
  \fi
  \vskip\belowcaptionskip}
\makeatother   

\begin{document}

\bibliographystyle{plain}
\maketitle
\newcommand{\ud}{\mathrm{d}}
\newcommand{\braket}[2]{\langle #1|#2\rangle}
\newcommand{\Bra}[1]{\left<#1\right|}
\newcommand{\Ket}[1]{\left|#1\right>}
\newcommand{\Braket}[2]{\left< #1 \right| #2 \right>}
\renewcommand{\th}{^\mathrm{th}}
\newcommand{\tr}{\mathrm{Tr}}
\newcommand{\id}{\mathds{1}}

\newtheorem{theorem}{Theorem}
\newtheorem{proposition}{Proposition}

\begin{abstract}
In topological quantum computation the geometric details of a particle
trajectory are irrelevant; only the topology matters. Taking this one
step further, we consider a model of computation that disregards even
the topology of the particle trajectory, and computes by permuting
particles. Whereas topological quantum computation requires anyons,
permutational quantum computation can be performed with ordinary
spin-1/2 particles, using a variant of the spin-network scheme of
Marzuoli and Rasetti. We do not know whether permutational computation
is universal. It may represent a new complexity class within
BQP. Nevertheless, permutational quantum computers can in
polynomial time approximate matrix elements of certain irreducible
representations of the symmetric group and simulate certain processes
in the Ponzano-Regge spin foam model of quantum gravity. No polynomial
time classical algorithms for these problems are known.
\end{abstract}

\section{Introduction}

There are now several models of quantum computation. These include
quantum circuits, topological quantum computation, adiabatic quantum
computation, quantum Turing machines, quantum walks, measurement-based
quantum computing, and the one clean qubit model. (See \cite{mythesis}
for an overview.) With the exception of the one clean qubit model, the
set of problems solvable in polynomial time in each of these models is
the same: BQP. This is proven by showing that each model can simulate
the others with only polynomial overhead\cite{Aharonov_adiabatic,
Freedman, Freedman2, Childs_univ, Raussendorf_Briegel,
Deutsch_networks}. Given these equivalences, one might ask why 
one should introduce new models of quantum computation. There are at
least three
reasons to do so. First, some models might be easier to 
physically implement than others. For example, the adiabatic model
seems particularly promising for implementation in superconducting
systems\cite{Kaminsky}, and the measurement based model seems
particularly promising
for optical implementation\cite{Nielsen_optical}. Second, new models
provide new conceptual frameworks for devising quantum algorithms. For
example, the topological model led directly to the discovery of
quantum algorithms for approximating Jones polynomials\cite{AJL}, and
the quantum walk model led to quantum algorithms for evaluating NAND
trees\cite{FGG}. Third, in rare instances, new models can lead to new
quantum complexity classes. The set of problems
solvable in polynomial time using the one clean qubit model is called
DQC1. It is believed that DQC1 contains some problems outside of P but
does not contain all of BQP \cite{Knill_DQC1}.

\begin{figure}
\begin{center}
\includegraphics[width=0.25\textwidth]{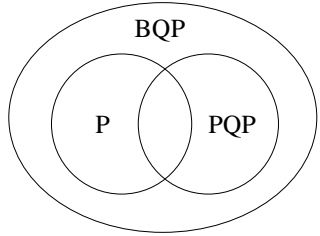}
\caption{\label{complexity} This diagram shows the conjectured
  relationships between classical polynomial time (P), quantum
  polynomial time (BQP), and permutational quantum polynomial time
  (PQP). Section \ref{inBQP} proves that PQP is contained in
  BQP. Sections \ref{mixed} and \ref{conclusion} give arguments for
  the conjecture that the containment is strict.}
\end{center}
\end{figure}

This paper considers the computational power obtained, not by braiding
and fusing anyonic particles in two dimensions, as is done in
topological quantum computation, but by permuting ordinary spins and
recoupling their angular momentum. The idea of
formulating a computational model based on spin recoupling was first
proposed by Marzuoli and Rasetti in \cite{Marzuoli}. Marzuoli and
Rasetti also suggested that their model\footnote{Their model differs
  slightly from the permutational model in that they allow continuous
  rotations in addition to the discrete operations of permutation and
  recoupling.} could be used to devise quantum algorithms for several
problems including the estimation of Ponzano-Regge partition
functions. Here we prove 
that permutational quantum computers can in polynomial time 
approximate matrix elements of irreducible representations of the
symmetric group in Young's orthogonal form, and we identify a class of
spin foams whose associated Ponzano-Regge amplitudes are efficiently
approximable on permutational quantum computers. We also
analyze fault tolerance of the permutational model, which seems
promising, as the computations are fully discrete, unaffected 
by perturbations to particle trajectory, and impervious to stray
magnetic fields provided they are uniform. Lastly, we prove that
permutational computers are efficiently simulatable by quantum circuits
and present evidence that the class of problems solvable with
polynomial resources on a permutational quantum computer, which I call
PQP, constitutes a new complexity class smaller than BQP but still
containing problems outside of P. Figure \ref{complexity} illustrates
the conjectured relations between PQP, P, and BQP. 
 
\section{The Model}
\label{model}

As is standard in quantum computing, we start by considering the
Hilbert space of $n$ two-level systems, such as spin-1/2
particles. It is conventional in quantum computing to use a basis for
this Hilbert space specified by the $\sigma_z$ Pauli operators on each
of the spins. These $n$ operators form a complete set of commuting 
observables. That is, each operator has eigenvalues $+1$ and $-1$, and
by specifying the simultaneous eigenvalues of all $n$ operators we
uniquely specify a state from an orthonormal basis of the
$2^n$-dimensional Hilbert space. In the context of quantum computation
this is called the computational basis. In the quantum circuit model,
one assumes the ability to construct pure basis states, and to make
projective measurements in the computational basis.

Identifying the $\sigma_z$ basis as the computational basis is not a
matter of pure convention. This basis consists of product
states. Preparation and measurement of such states can be done in
principle using physically realistic operations that do not require
many-body interactions. In contrast, there exist globally entangled
states of $n$ spins that require exponentially many local operations
(\emph{e.g.} quantum gates) to construct, and entangled measurements that
require exponentially many local operations to perform. Nevertheless,
there exist other choices for a complete set of commuting observables
such that state preparation and measurement remain physically
plausible. In particular, there exist exponentially many complete sets
of commuting observables constructable from total spin angular
momentum operators.

For any $i \in \{1,2,\ldots,n\}$, let
\begin{equation}
\label{vecS}
\vec{S}_i = \frac{1}{2} \left( \begin{array}{c}
\sigma_x^{(i)} \\
\sigma_y^{(i)} \\
\sigma_z^{(i)}
\end{array}
\right)
\end{equation}
be the spin angular momentum operator for the $i\th$ spin-1/2
particle. Similarly for any $a \subseteq \{1,2,\ldots,n\}$, let
\[
S^2_a = \left( \sum_{i \in a} \vec{S}_i \right) \cdot \left( \sum_{i
    \in a} \vec{S}_i \right)
\]
be the total spin angular momentum operator for the set $a$ of
spins. (Here $\cdot$ indicates the three-dimensional dot product.) If
$a$ and $b$ are disjoint sets or if one is a subset of the other
then $S^2_a$ commutes with $S^2_b$.  

The total angular momentum operators corresponding to certain sets of
subsets of spins, together with a total azimuthal angular momentum
operator form a complete set of commuting observables. For example,
consider the case of three spin-1/2 particles. The following is one
complete set of commuting observables.
\begin{eqnarray}
\label{firstchoice}
S_{\{123\}}^2 & = & \left( \vec{S}_1 + \vec{S}_2 + \vec{S}_3 \right)^2
\nonumber \\
S_{\{12\}}^2 & = & \left( \vec{S}_1 + \vec{S}_2 \right)^2 \\
Z_{\{123\}} & = & \frac{1}{2} \left( \sigma_z^{(1)} + \sigma_z^{(2)} +
  \sigma_z^{(3)} \right) \nonumber
\end{eqnarray}
Here is another complete set of commuting observables. 
\begin{eqnarray}
\label{secondchoice}
S_{\{123\}}^2 & = & \left( \vec{S}_1 + \vec{S}_2 + \vec{S}_3 \right)^2
\nonumber \\
S_{\{23\}}^2 & = & \left( \vec{S}_2 + \vec{S}_3 \right)^2 \\
Z_{\{123\}} & = & \frac{1}{2} \left( \sigma_z^{(1)} + \sigma_z^{(2)} +
  \sigma_z^{(3)} \right) \nonumber
\end{eqnarray}
Diagramatically, we can represent these two choices by binary trees
\begin{center}
\includegraphics[width=0.3\textwidth]{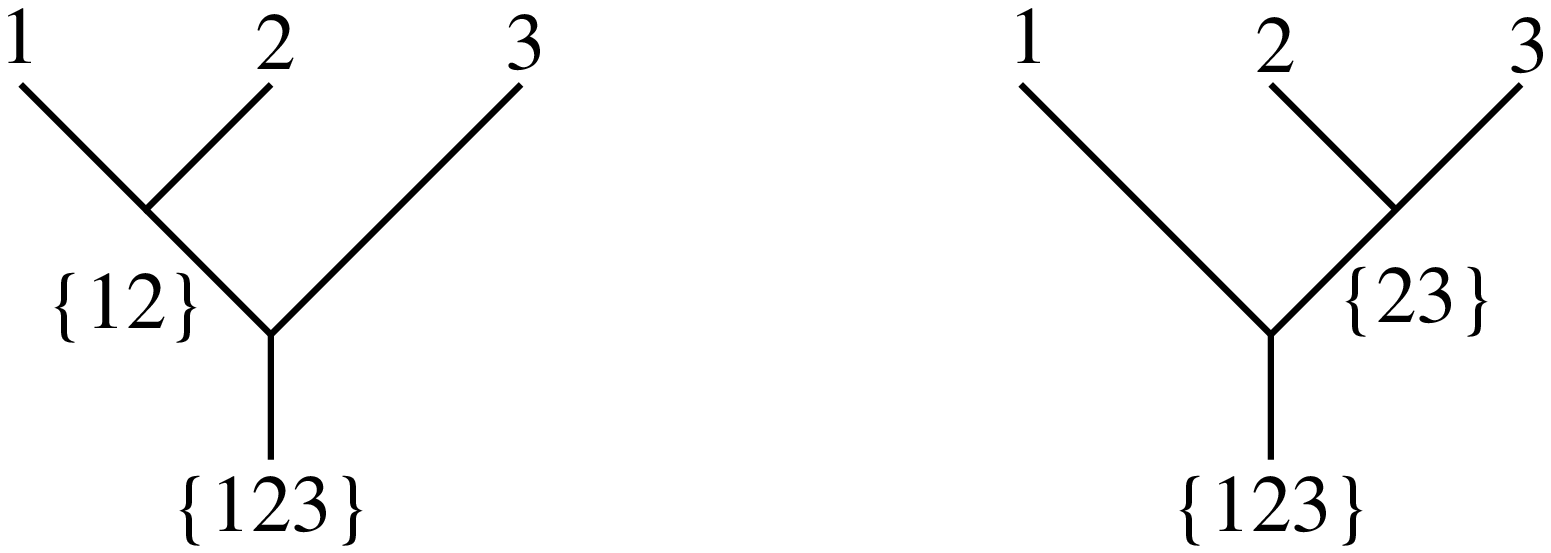}
\end{center}
At each trivalent node of the tree we have two incoming edges
corresponding to operators $S^2_a$ and $S^2_b$ for two sets of spins
$a$ and $b$, and one outgoing edge corresponding to the operator
$S^2_{a \cup b}$. The possible eigenvalues of these operators are
given by the following standard rules for angular momentum addition in quantum
mechanics\cite{Sakurai}. For any set of spins $a$, the
allowed eigenvalues of $S^2_a$ are of the form $j_a (j_a + 1)$ where
$j_a$ is a nonnegative integer or half-integer. The
number $j_a$ is referred to as the total angular momentum of the set
of spins $a$. A single spin with total angular momentum $j$ is
referred to as a spin-$j$ particle. The possible eigenvalues of  
$S^2_{a \cup b}$ are subject to the constraints 
\[
j_{a \cup b} + j_a + j_b \in \mathbb{Z}
\]
and
\[
|j_a-j_b| \leq j_{a \cup b} \leq j_a + j_b.
\]

For a given complete set of commuting angular momentum observables, we can
diagrammatically denote the corresponding basis states by labeling
each edge of the tree with the total angular momentum $j$ for the
corresponding subset of spins. For example, choice (\ref{firstchoice})
yields the following basis for the eight-dimensional Hilbert space of
three spin-1/2 particles.
\begin{center}
\includegraphics[width=0.6\textwidth]{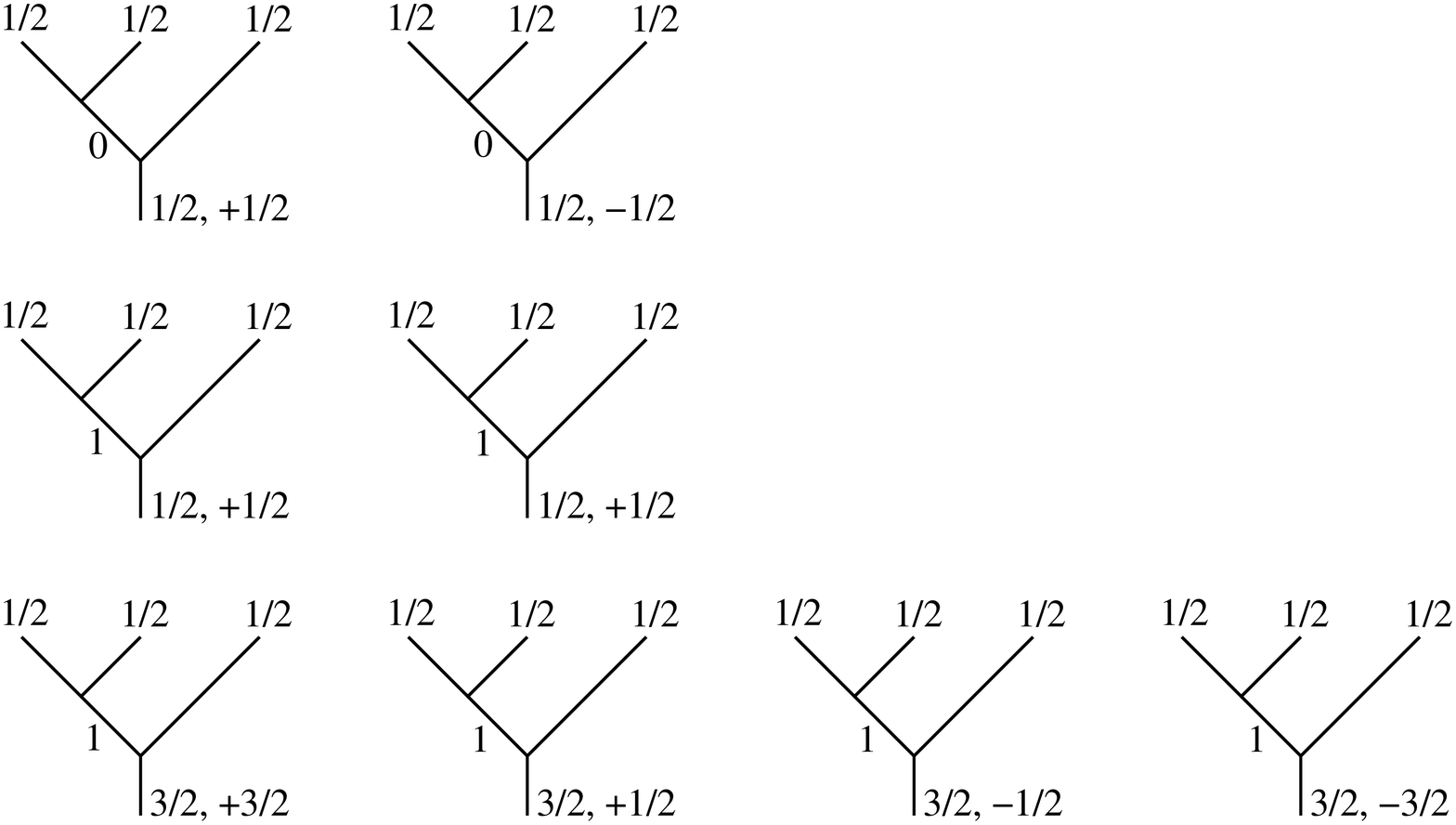}
\end{center}
The extra label on the root of the tree indicates the eigenvalue of 
azimuthal angular momentum operator $Z_{\{123\}}$. 

This idea generalizes straightforwardly to any number of spins. For
example, the binary trees for four spins are shown below.
\begin{center}
\includegraphics[width=0.8\textwidth]{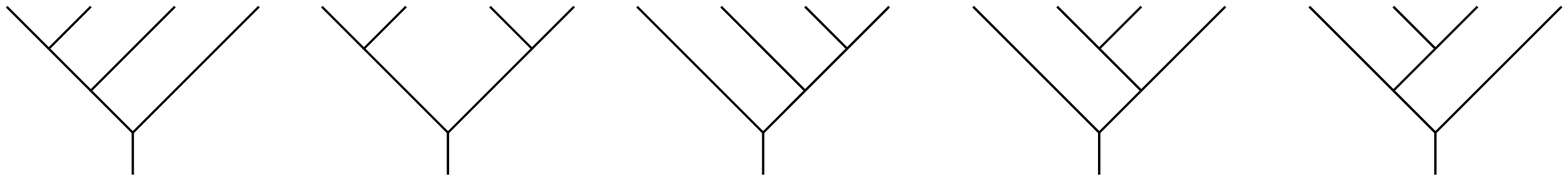}
\end{center}
These trees correspond to five different orthonormal bases for the
16-dimensional Hilbert space for four spin-1/2 particles. It is not
hard to see that the number of binary trees on $n$ spins scales
exponentially with $n$.

In the most basic permutational model of quantum computation we have
$n$ spins, and can prepare any state corresponding to a labeled
binary tree of $n$ leaves. After preparing a state we apply an
arbitrary permutation to the $n$ particles. Lastly, we measure any
complete commuting set of total angular momentum  operators, thereby
performing a projective measurement in an orthonormal basis
corresponding to one of the unlabeled binary trees of $n$
leaves. By repeating this process, we can sample from the probability
distribution defined by this measurement. Thus we can estimate the
probability corresponding to a particular final state (labeled
binary tree) to within $\pm \epsilon$ using $O(1/\epsilon^2)$ trials.

It seems mathematically natural to define a stronger version of the
permutational model using the amplitudes rather than the
probabilities. That is, let $\pi \in S_n$, and let $\lambda,\lambda'$ 
be any pair of labeled binary trees of $n$ leaves. We call the
corresponding states of $n$ spin-$1/2$ particles $\ket{\lambda}$ and
$\ket{\lambda'}$. Let $U_\pi$ be the transformation induced by
permuting these $n$ spins according to $\pi$:
\[
U_{\pi} \ket{z_1} \otimes \ket{z_2} \otimes \ldots \otimes \ket{z_n} =
\ket{z_{\pi(1)}} \otimes \ket{z_{\pi(2)}} \otimes \ldots \otimes
\ket{z_{\pi(n)}}.
\]
In the strong permutational model we assume that we can perform an
experiment to determine the real and imaginary parts of the amplitude
$\bra{\lambda} U_\pi \ket{\lambda'}$
to precision $\pm \epsilon$ in $\mathrm{poly}(1/\epsilon)$
time. Physically, such an experiment may be harder to perform than that of
the basic permutational model, although it could be done in principle using
coherently controlled state preparations and interferometric
measurements such as the Hadamard test. (\emph{cf.} \cite{AJL}, section
2.2) On the other hand, the strong permutational model seems to be
more convenient for defining a new complexity class and formulating
new quantum algorithms. In this paper, we analyze the strong
permutational model. Performing the analogous analysis in the weak
model is a straightforward exercise.

For a set $a$ of $n$ spins, the operator $Z_a$ commutes with all
permutation operators $\{U_\pi|\pi \in S_n\}$ and all total angular momentum
operators $\{S_b|b \subseteq a\}$. Thus,
the permutational model amplitudes can be factored as $\bra{\lambda}
U_\pi \ket{\lambda'} = \delta_{m,m'} f(\pi,S^2_a,S^2_b,\ldots)$, where
$m$ and $m'$ are the eigenvalues $Z_a \ket{\lambda} = m \ket{\lambda}$
and $Z_a \ket{\lambda'} = m' \ket{\lambda'}$, and $f$ is a function
only of the permutation $\pi$ and the total angular momenta of the
various subsets $a,b,\ldots$. Therefore, we henceforth describe all
permutational computations in terms of amplitudes $\bra{\lambda} U_\pi
\ket{\lambda'}$, where $\lambda$ and $\lambda'$ are binary trees
labeled only with $j$ values. It is implicit that $m = m'$. Beyond
that we do not care about the actual values of $m$ and $m'$.

In the quantum circuit model, we measure the length of a computation
by the number of elementary quantum gates. At first glance it seems
natural to seek some analogous measure of length for permutational
computations. One choice would be to imagine the $n$ spins arranged
along a line, and consider the transposition of a pair of neighbors as
an elementary operation. Such transpositions generate the symmetric
group, thus any permutational computation could be built up from these
steps. However, given any permutation in $S_n$ it
is easy to find a sequence of at most $O(n^2)$ transpositions to
implement it. In fact, the well-known bubblesort algorithm can be
viewed as a method for finding such a sequence. Thus, all of the
possible permutational computations on $n$ spins can be achieved in
$\mathrm{poly}(n)$ time. Hence we can ignore computation length and
simply define PQP to be, roughly speaking, the set of problems
solvable by estimating amplitudes of the form $\bra{\lambda} U_\pi
\ket{\lambda'}$ on polynomially many spins, to polynomial precision.

To make a completely precise definition of PQP we must specify what
sort of computer we use to control the experiment. That is, the
computer is given a problem instance, and based on that it decides 
which amplitudes of the form $\bra{\lambda} U_\pi \ket{\lambda'}$ to
estimate. It then transmits instructions to the experimental
apparatus, and receives the measurement outcomes, which it
postprocesses in order to answer the problem. If we choose a P machine then
PQP trivially contains P. To allow a more meaningful comparison
between the permutational model and classical polynomial time
computation we therefore use a logspace machine. We can abstractly
define PQP to be the set of problems solvable by a logspace machine
with access to an oracle that provides amplitudes $\bra{\lambda} U_\pi
\ket{\lambda'}$ for polynomially many spins, to polynomial
precision. Thus PQP trivially contains the complexity class L, but
whether PQP contains P remains an interesting open
question\footnote{Choosing an NC1 machine rather than a logspace
  machine is also reasonable\cite{Shor_Jordan}. The results obtained
  in this paper all hold for either choice. See \cite{complexity_zoo}
  for definitions of L and NC1.}.

\section{PQP is contained in BQP}
\label{inBQP}

Permutational quantum computation can be analyzed either using the
computational basis or using a basis of $j$-labeled trees arising
from total angular momentum operators. Either method of
analysis yields fairly directly a proof that $\mathrm{PQP} \subseteq
\mathrm{BQP}$. Throughout this paper we exclusively use the
basis of $j$-labeled trees. This basis makes the connection to anyonic
computation clearer, as the bases used to analyze anyonic quantum
computation are $q$-deformations of these (see \cite{Kauffman}).

Let $\lambda$ and $\lambda'$ be a pair of $j$-labeled binary trees
with $n$ leaves, and let $\pi$ be some permutation in $S_n$. Any
amplitude of the form $\bra{\lambda} U_\pi \ket{\lambda'}$ can be
calculated using the following two diagrammatic rules.
\begin{equation}
\label{recouple}
\includegraphics[width=0.5\textwidth]{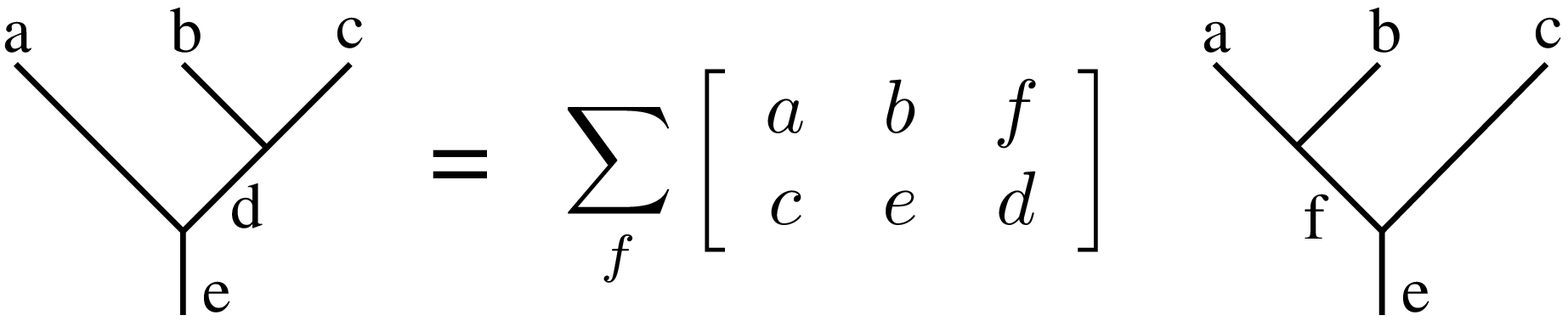} \\
\end{equation}

\begin{equation}
\label{twist}
\includegraphics[width=0.35\textwidth]{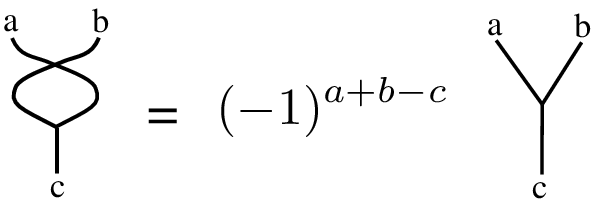}
\end{equation}

\begin{figure}
\begin{center}
\includegraphics[width=0.6\textwidth]{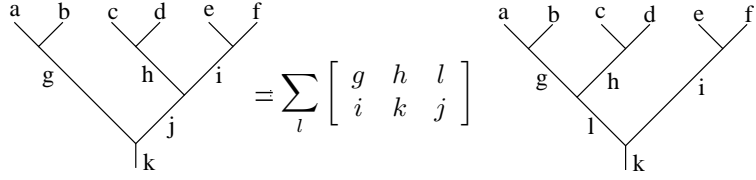}
\caption{\label{subtree} Here we apply the recoupling rule
  \ref{recouple} to an internal node of a tree. Rules \ref{recouple}
  and \ref{twist} apply whether the $j$ labels refer to the angular momenta of
  individual particles or sets of particles.}
\end{center}
\end{figure}

Rule \ref{recouple} is a change of basis between the simultaneous
eigenbasis of $S_{12}$ and $S_{123}$ and the simultaneous eigenbasis
of $S_{23}$ and $S_{123}$. Furthermore, if instead of three spins
$1,2,3$ we have three sets of spins $a_1,a_2,a_3$, then the same
formula rule converts between the simultaneous eigenbasis of $S_{a_1
  \cup a_2}$ and $S_{a_1 \cup a_2 \cup a_3}$ and the simultaneous
eigenbasis of $S_{a_2 \cup a_3}$ and $S_{a_1 \cup a_2 \cup a_3}$. In
other words, we can apply this diagrammatic rule to any internal node
of a $j$-labeled binary tree, as illustrated in figure
\ref{subtree}. As discussed in \cite{Marzuoli}, the recoupling tensor
is
\begin{equation}
\label{recoupling_tensor}
\left[ \begin{array}{ccc} a & b & f \\
c & e & d \end{array} \right] = (-1)^{a+b+c+e} \sqrt{(2d+1)(2f+1)}
\left\{ \begin{array}{ccc} a & b & f \\
c & e & d \end{array} \right\},
\end{equation}
where $\left\{ \begin{array}{ccc} a & b & f \\ c & e & d \end{array}
\right\}$ is the $6j$ symbol for $SU(2)$. The $6j$ symbol can be
calculated using the Racah formula\cite{mathworld}
\begin{equation}
\label{racah}
\left\{ \begin{array}{ccc} a & b & f \\ c & e & d \end{array}
\right\} = \sqrt{\Delta(a,b,f) \Delta(a,e,d) \Delta(c,b,d)
  \Delta(c,e,f)} \sum_t \frac{(-1)^t (t+1)!}{f(t)},
\end{equation}
where
\[
\Delta(a,b,c) = \frac{(a+b-c)! (a-b+c)! (-a+b+c)!}{(a+b+c+1)!}
\]
and
\[
f(t) = 
(t-a-b-f)!(t-a-e-d)!(t-c-b-d)!(t-c-e-f)!(a+b+c+e-t)!(b+f+e+d-t)!(f+a+d+c-t)!.
\]
The sum in equation \ref{racah} is over all $t$ such that the
factorials in $f(t)$ all have nonnegative arguments. The main thing to
notice about these formulas is that the recoupling tensor 
$\left[  \begin{array}{ccc} a & b & f \\ c & e & d \end{array} \right]$
can be computed in polynomial time provided that $a,b,c,d,e,f$ are all
at most polynomially large.

To illustrate these rules we first work out an example by
hand. Suppose we have three spin-1/2 particles and we wish to compute
the amplitude
\[
\includegraphics[width=0.28\textwidth]{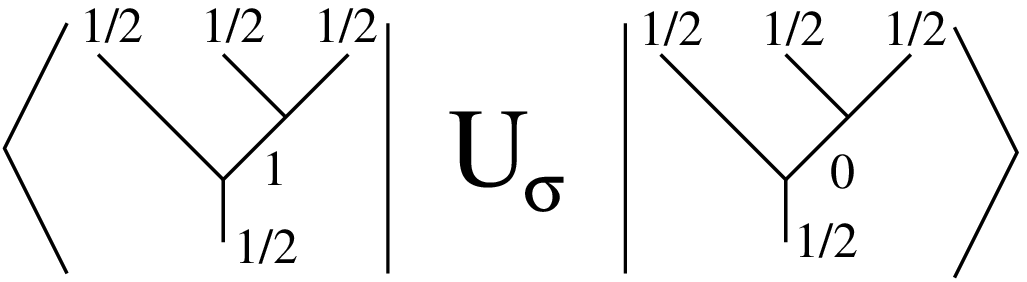},
\]
where $\sigma$ is the permutation that swaps the leftmost pair of
spins. We can rewrite this as
\begin{equation}
\label{twistree}
\includegraphics[width=0.24\textwidth]{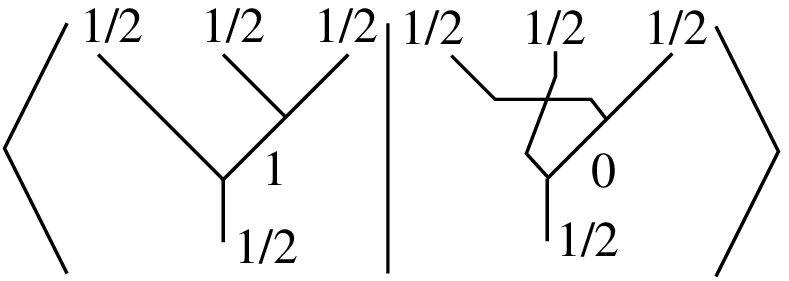}.
\end{equation}
Because the crossed sub-branches do not come from the same branch we
cannot at this point apply rule \ref{twist}. Instead, we must first
apply rule \ref{recouple}, obtaining
\begin{equation}
\label{withj}
\includegraphics[width=0.4\textwidth]{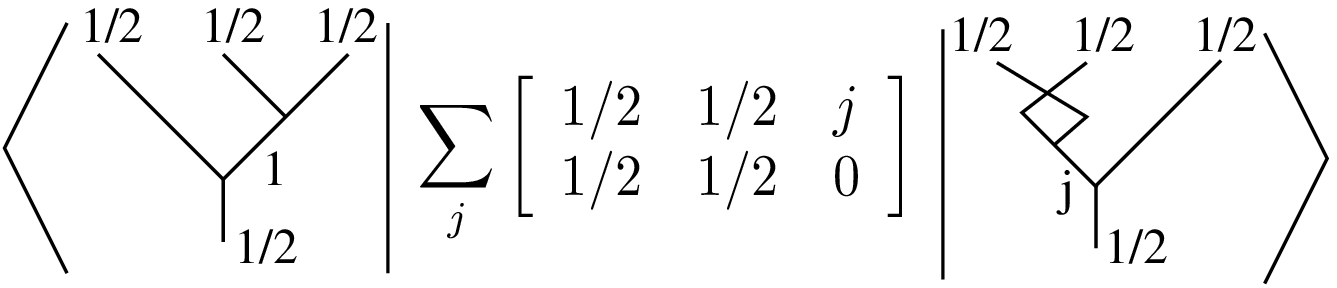}.
\end{equation}
Applying rule \ref{twist} then yields
\[
\includegraphics[width=0.48\textwidth]{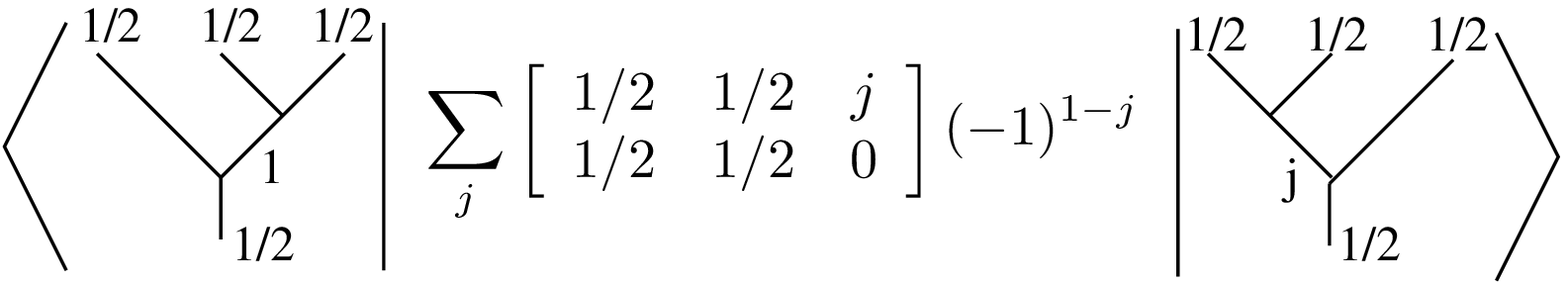}.
\]
Then we can apply rule \ref{recouple} again to bring the two trees
into the same form.
\begin{equation}
\label{sameform}
\includegraphics[width=0.66\textwidth]{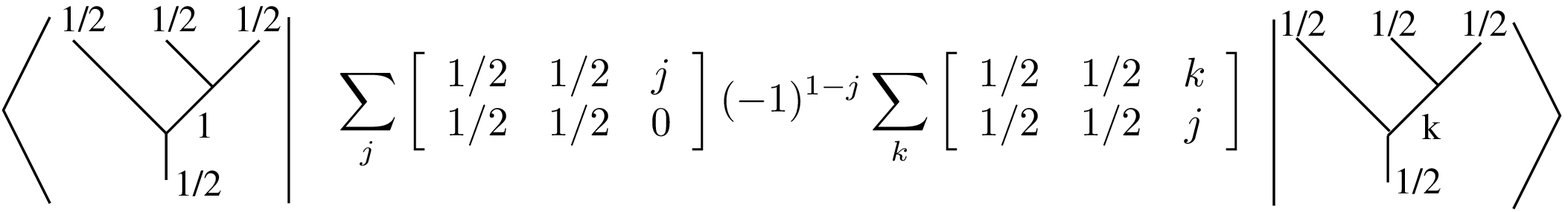}
\end{equation}
The total angular momentum operators are Hermitian. Thus, their
eigenstates form an orthonormal basis. In other words, distinct
labelings of a given binary tree correspond to orthonormal
states. Thus,
\[
\includegraphics[width=0.28\textwidth]{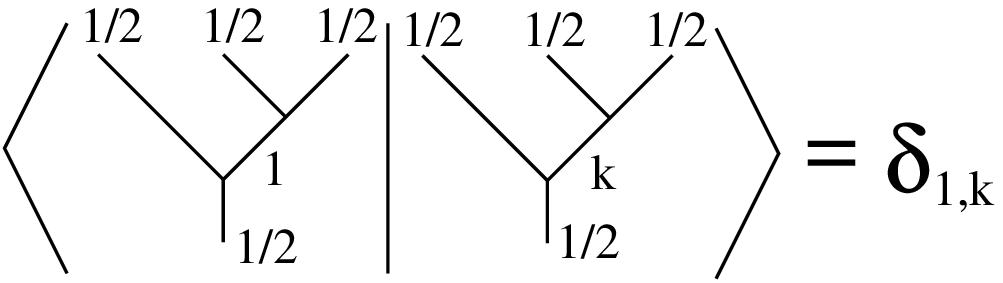}
\]
so the expression \ref{sameform} evaluates to
\[
\sum_j \left[ \begin{array}{ccc} 1/2 & 1/2 & j \\
1/2 & 1/2 & 0 \end{array} \right] (-1)^{1-j} 
\left[ \begin{array}{ccc} 1/2 & 1/2 & 1 \\
1/2 & 1/2 & j \end{array} \right].
\]
The sum over $j$ is in principle over all integers and
half-integers. However, the recoupling tensors are only nonzero in a
finite set of cases. Specifically, in any nonzero term, the $j$-labels
on a tree must obey the laws of angular momentum addition:
\begin{equation}
\label{laws}
\includegraphics[width=0.33\textwidth]{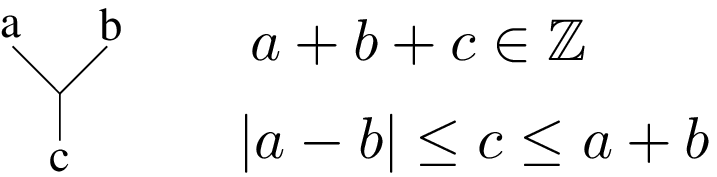}
\end{equation}
Thus, recalling expression \ref{withj}, we see that the sum is over
$j=0$ and $j=1$, and evaluates to
\[
\includegraphics[width=0.95\textwidth]{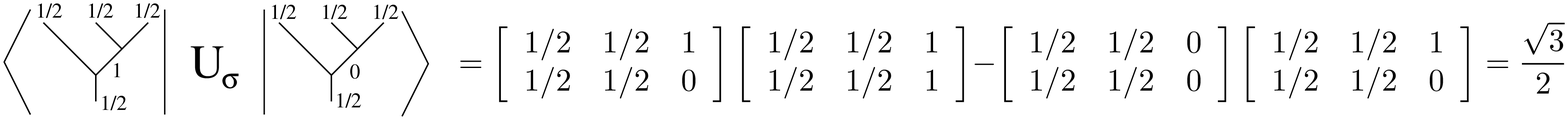}.
\]
This example illustrates all of the principles needed for the general
case. Given $\pi \in S_n$ and two $j$-labeled binary trees $\lambda$
and $\lambda'$ of $n$ leaves, we can compute the amplitude
$\bra{\lambda'} U_\pi \ket{\lambda}$ by first applying $\pi$ to the
leaves of $\lambda$, obtaining a twisted tree $\tilde{\lambda}$, as
in expression \ref{twistree}. Then, we apply a sequence of
recoupling (\ref{recouple}) and twist (\ref{twist}) moves to
untangle the tree, leaving a superposition over labelings of an
ordinary tree. Then, we use some sequence of recoupling moves to
bring this tree into the same form as $\lambda'$. Then we apply the
orthonormality of different labelings of the tree.

The quantum circuit algorithm for approximating $\bra{\lambda'} U_\pi
\ket{\lambda}$ essentially mirrors this process. Given
$\lambda$ and $\lambda'$ with $n$ leaves and $\pi \in S_n$, it is
clear that in $\mathrm{poly}(n)$ time on a classical computer we can
compute a sequence of polynomially many twists and recouplings that
untangles $\tilde{\lambda}$ and then brings it into the same form
as $\lambda'$. Thus, the task to perform on the quantum computer is
to track the resulting superposition over labelings. Generically,
the number of terms in this superposition grows exponentially in the
number of recoupling moves performed. On a quantum computer we build
such superpositions by constructing quantum circuits to implement the
unitary transformations of rules \ref{twist} and \ref{recouple}.

In a $j$-labeled tree with $n$ spin-1/2 leaves, every $j$-label must
come from the set $\{0,1/2,1,\ldots,n/2\}$. This is a consequence of
the condition $c \leq a+b$ from \ref{laws}. We can
therefore use $\lceil \log_2 (n+1) \rceil$ qubits to store each
label. We need not use a quantum register to track the shape of the
tree, only its labeling. This is because, by starting with a given
tree $\lambda$ and applying a sequence of recoupling and twist moves
we only obtain superpositions over different labelings of a given
tree, never superpositions over different trees. Furthermore, we need
not track the labels on the leaves or the root, as these are left
invariant by all twist and recoupling moves. 

Let's return to the example evaluated by hand above. Discarding the
root and leaves as fixed, we need only track one $j$-value. By the
general argument above, this $j$ value must lie in the set
$\{0,1/2,1,3/2\}$. We correspondingly use a register of two qubits to
encode the value of $j$. (In fact, in this particular case, $j$ can only take
on values 0 and 1, but we shall ignore this extra information.) To
represent the initial state $\lambda$ we initialize a register of two
qubits to the state $\ket{0}$. Then we apply the recoupling and twist
moves in sequence, which in this context are unitary transformations
on the two-qubit register.
\begin{eqnarray*}
\ket{0} & \to & \sum_j \left[ \begin{array}{ccc} 1/2 & 1/2 & j \\
1/2 & 1/2 & 0 \end{array} \right] \ket{j} \\
 & \to & \sum_j \left[ \begin{array}{ccc} 1/2 & 1/2 & j \\
1/2 & 1/2 & 0 \end{array} \right] (-1)^{1-j} \ket{j} \\
& \to & \sum_j \left[ \begin{array}{ccc} 1/2 & 1/2 & j \\
1/2 & 1/2 & 0 \end{array} \right] (-1)^{1-j} \sum_k \left[
\begin{array}{ccc} 1/2 & 1/2 & k \\ 1/2 & 1/2 & j \end{array} \right] \ket{k}
\end{eqnarray*}
We then use the Hadamard test to approximate the real and imaginary
parts of the amplitude associated with $k=1$.

The general case is a straightforward extension of this example. All
that remains is to show that quantum circuits can efficiently
implement the unitary transformations corresponding to the twist and recoupling
moves. The unitary transformation of the twist (eq. \ref{twist}) is simply a
diagonal unitary with $+1$ and $-1$ entries along the
diagonal. Furthermore, for a given set of $j$ labels it is easy to
classically compute whether the corresponding sign should be $+1$ or
$-1$. Therefore, it can be implemented using polynomially many quantum
gates via a a standard technique called phase
kickback\cite{kickback}. The recoupling transformation acts on only
six registers of $\lceil \log_2 (n+1) \rceil$ qubits
each. Furthermore, by the Racah formula, the matrix elements of the
transformation induced on these qubits can all be efficiently computed
classically. As discussed in section 4.5 of \cite{Nielsen_Chuang}, any unitary
transformation on logarithmically many qubits with efficiently
computable matrix elements can be implemented using polynomially many
quantum gates\footnote{Alternatively, we need only observe that the
  twist and recoupling tensors are sparse and have efficiently
  computable matrix elements. They are therefore implementable using
  the general construction of \cite{sparse}.}.

\section{Approximating Irreps of the Symmetric Group}
\label{irreps}

In this section, we show that a permutational quantum computer can, in
polynomial time, approximate matrix elements of certain irreducible
representations of the symmetric group. The problem of computing
explicit irreducible representations of the symmetric group has been
studied classically\cite{Hamermesh, Boerner, Wu_Zhang1, Wu_Zhang2,
  Clifton, Rettrup, Pauncz}, and no polynomial time algorithm is
known. Thus, this result provides some evidence that PQP is not
contained in P. Furthermore, it provides a potentially useful
application for permutational quantum computers should one ever be
built. Lastly, as shown in the preceding section, PQP is contained in
BQP, thus any quantum algorithm for permutational quantum computers is
also automatically a quantum algorithm for standard quantum
computers. An efficient algorithm for approximating matrix elements of
irreducible representations of the symmetric group on standard quantum
computers was derived from a somewhat different point of view in
\cite{Jordan_groups}.

Let $a$ be a set of $n$ spin-1/2 particles, and let $\mathcal{H}_a$ be
the corresponding $2^n$-dimensional Hilbert space. The symmetric group
$S_n$ acts on $\mathcal{H}_a$ in a straightforward way. For
any $\pi \in S_n$, we have the action
\[
U_{\pi} \ket{z_1} \otimes \ket{z_2} \otimes \ldots \otimes \ket{z_n} =
\ket{z_{\pi(1)}} \otimes \ket{z_{\pi(2)}} \otimes \ldots \otimes
\ket{z_{\pi(n)}}.
\]
The map $\pi \to U_\pi$ is a homomorphism from $S_n$ to the unitary
group $U(2^n)$. In other words, it is a unitary representation of
$S_n$. This representation is reducible. For all $\pi \in S_n$,
$U_\pi$ commutes with the total angular momentum operator
$S^2_{a}$ and the total $Z$-angular-momentum operator $Z_a$. Thus, the
simultaneous eigenspaces of $S^2_a$ and $Z_a$ are each invariant under
the action of $S_n$. The action of $S_n$ on any of these eigenspaces
is therefore a representation of $S_n$. These representations are all
irreducible\cite{Pauncz2}.

\begin{figure}
\begin{center}
\includegraphics[width=0.45\textwidth]{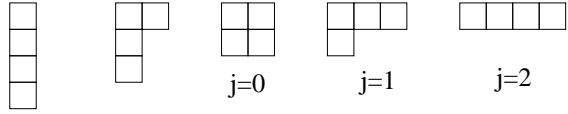}
\caption{\label{Young} Here we list all the Young diagrams of four
  boxes. These correspond to all of the irreducible representations of
  $S_4$. Let $a$ be a set of four spin-1/2 particles. The Young
  diagrams with two rows correspond to irreducible representations
  arising from the action of $S_4$ on the angular momentum eigenspace
  $\{\ket{\psi}:\ S^2_a \ket{\psi} = j(j+1) 
   \ket{\psi}\ \mathrm{and}\ Z_a \ket{\psi} = m \ket{\psi}\}$
  for any allowable $m$.}
\end{center}
\end{figure}

The irreducible representations of $S_n$ are usually specified by
Young diagrams. A Young diagram for $S_n$ is a partition of $n$ boxes
into rows, such that no row is longer than the row above it, as
illustrated in figure \ref{Young}. Let $\mathcal{V}_j$ be the
eigenspace of $S^2_a$ with eigenvalue $j(j+1)$. Within any fixed eigenspace
of $Z_a$, the action of $S_n$ on $\mathcal{V}_j$ is the irreducible
representation whose Young diagram has two rows, where the overhang of
the top row over the bottom is $2j$ \cite{Pauncz2}.

To obtain an explicit matrix representation of $S_n$ we must choose a
basis. We can choose a basis for the eigenspaces of $S^2_a$ and $Z_a$ by
finding subsets $b,c,\ldots$ of $a$ such that $S^2_a$ and $Z_a$ together
with $S^2_b, S^2_c, \ldots$ form a complete a set of commuting
observables. As described in section \ref{model}, the possible choices
for subsets $c,d,\ldots$ correspond bijectively to the rooted binary
trees of $n$ leaves. The different trees give us different
bases. Within a given basis, the different basis states correspond to
different $j$-labelings of the chosen tree.

The binary tree bases have the special property of being subgroup
adapted. Let $G$ be a group and let $H$ be a subgroup of $G$. Any
representation $\rho_G$ of $G$ (homomorphism from $G$ to a group of linear
transformations) yields a representation $\rho_H$ of $H$ if we simply
restrict its domain to $H$. However, an irreducible representation of
$G$ does not necessarily remain irreducible when restricted to $H$. In
this case $\rho_H$ is isomorphic to a direct sum of irreducible
representations of $H$. Suppose we choose a basis for the
representation. Now $\rho_G$ becomes a map from group elements to
matrices. The basis is adapted for the subgroup $H$ if $\rho_G$ maps
the elements of $H$ to matrix direct sums of irreducible
representations of $H$. As discussed elsewhere\cite{Jordan_groups,
  generalft}, subgroup adapted bases are very useful in mathematics,
physics, and both quantum and classical computing.

Recalling that the representation of $S_n$ corresponding to any fixed
label on the root is irreducible, and examining the diagrammatic rules
\ref{twist} and \ref{recouple} we see that the binary tree basis is
subgroup adapted for each of the subgroups of $S_n$ that fix all of
the leaves other than those on a given subtree. In this example, 
\[
\includegraphics[width=0.35\textwidth]{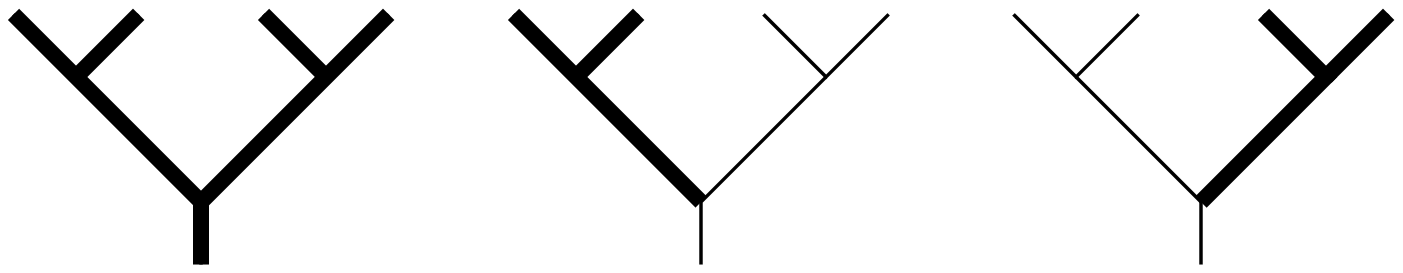}
\]
the left and right subtrees shown indicate that the basis is
adapted to two subgroups of $S_4$, each isomorphic to $S_2$. Whereas,
in this example,
\[
\includegraphics[width=0.35\textwidth]{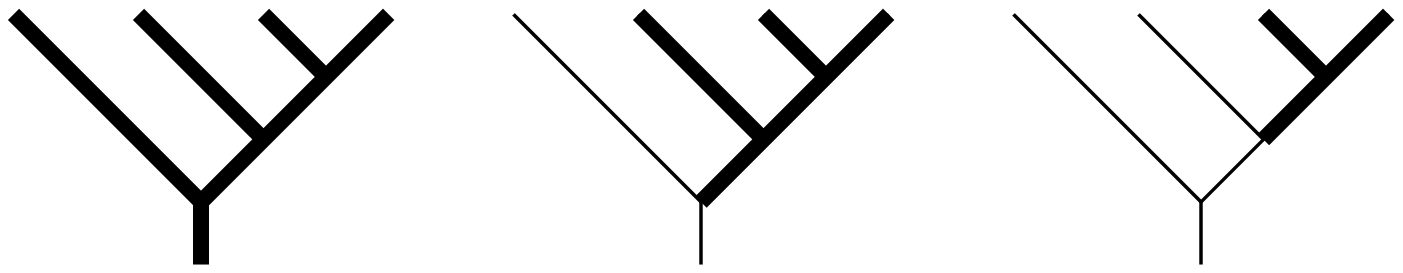}
\]
the subtrees shown indicate that the basis is adapted 
to two subgroups of $S_4$, one isomorphic to $S_2$ and one isomorphic
to $S_3$.

The most standard and widely used basis for unitary representations of
the symmetric group is the Young-Yamanouchi basis. The resulting maps
from permutations to matrices is often called Young's orthogonal
form (introduced in 1927 by Alfred Young\cite{Young}). Suppose
that the elements of $S_n$ act by permuting a set of
objects arranged along a line. The set of permutations in $S_n$ that
leaves all but the rightmost $m$ objects untouched is a subgroup of
$S_n$ isomorphic to $S_m$. Young's orthogonal form is adapted to this
chain of subgroups isomorphic to $S_n \supset S_{n-1} \supset \ldots
\supset S_3 \supset S_2$.

Using angular momentum operators we can construct a basis adapted to
this same chain of subgroups. The corresponding binary trees are those
of the following form.
\begin{equation}
\label{centipedes}
\includegraphics[width=0.45\textwidth]{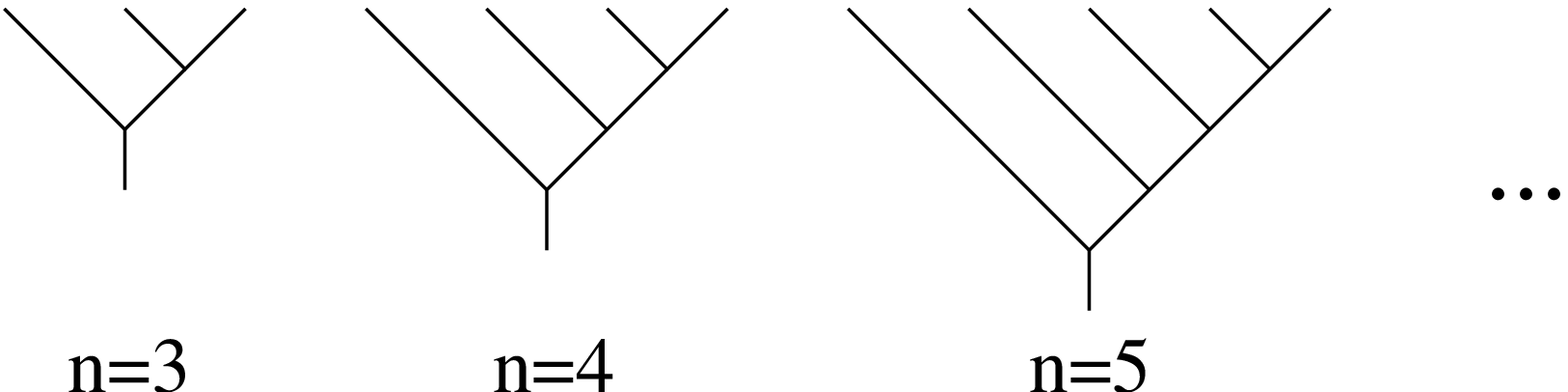}
\end{equation}
This might lead one to guess that the matrix representation of $S_n$
arising from this type of binary tree is identical to Young's
orthogonal form. Indeed, this is correct, as is shown, for example, in
\cite{Pauncz2}.

By the above discussion, permutational quantum computers can
efficiently approximate matrix elements from Young's orthogonal
form. To state this result more precisely, we must describe how the
problem instance is input to the computer. Let $\lambda$ be a
Young diagram of $n$ boxes. By labeling these boxes from 1 to $n$ such
that the numbers in any column are increasing downward, and the
numbers in any row are increasing rightward, we obtain a standard
Young tableau of shape $\lambda$. For example, the standard Young
tableaux of shape
\[
\includegraphics[width=0.043\textwidth]{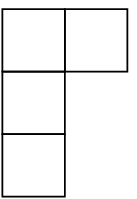}
\]
are
\[
\includegraphics[width=0.2\textwidth]{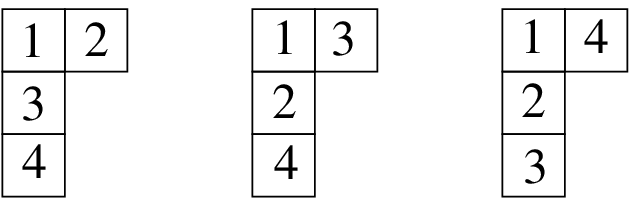}.
\]
For a Young diagram $\lambda$, the corresponding irreducible
representation $\rho_\lambda$ in Young's orthogonal form can be
formulated as a linear transformation on the formal span of all
standard Young tableaux of shape $\lambda$. Thus standard Young
tableaux of shape $\lambda$ index the rows and columns of the matrices
of representation $\rho_\lambda$.

\begin{figure}
\begin{center}
\includegraphics[width=0.8\textwidth]{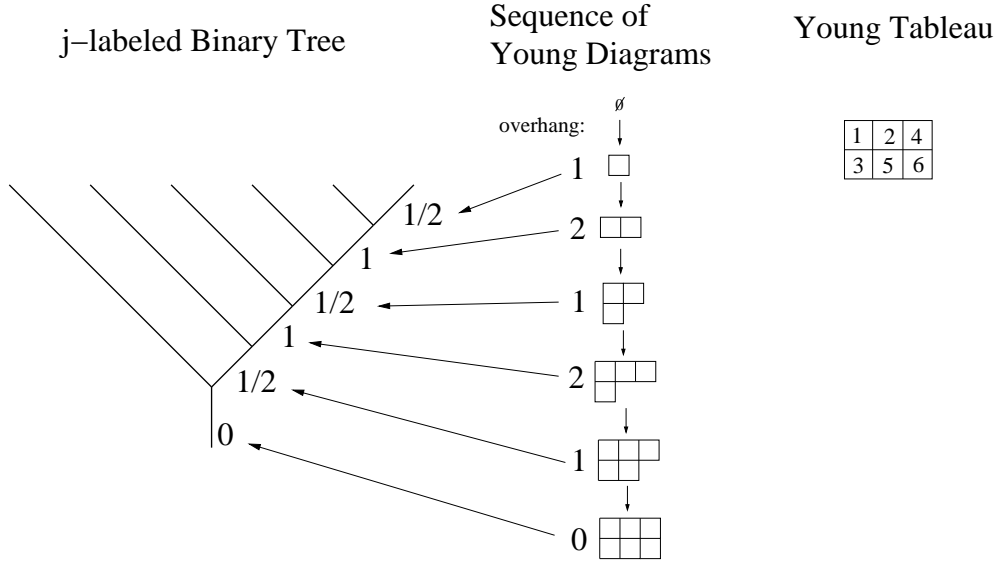}
\caption{\label{correspondence} Each Young tableau corresponds to a
  sequence of Young diagrams. If the Young tableau has only two rows,
  then these diagrams can be characterized by the overhang of the top
  row over the bottom row. Dividing by two, this sequence of overhangs
  then corresponds a $j$-labeling of a binary tree, as shown. We have
  omitted the labels from the leaves, as they are all spin-$1/2$.}
\end{center}
\end{figure}

If $\lambda$ has two rows, then the standard Young tableaux of shape
$\lambda$ correspond bijectively to the $j$-labelings of a binary tree as
follows. We can think of the numbering of boxes in a standard Young
tableau as an instruction for building the final Young diagram by
adding one box at a time. For example, the Young tableau
\[
\includegraphics[width=0.07\textwidth]{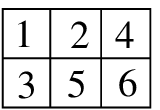}
\]
corresponds to the sequence
\begin{equation}
\label{build}
\includegraphics[width=0.6\textwidth]{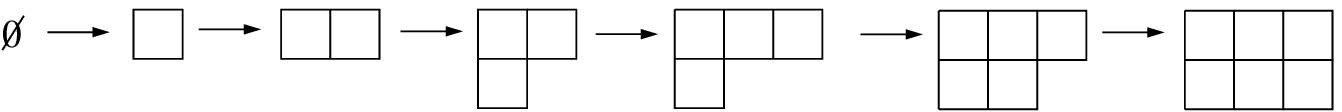}.
\end{equation}
The condition that numbers increase downward in each column and
rightward in each row is equivalent to the condition that the
configuration of boxes after each step is a valid Young diagram. To
each Young diagram in the sequence we can associate an ``overhang'',
the number of boxes in the top row minus the number of boxes in the
second row. The labeled tree corresponding to a given Young tableau is
of the type shown in diagram \ref{centipedes}, and the sequence of
overhangs, each divided by two, labels the edges from top to bottom
down the right hand side. Figure \ref{correspondence} gives an
example.

A permutational quantum computer can solve the following problem in
time $\mathrm{poly}(n, 1/\epsilon)$.\\ \\
\noindent
\begin{minipage}[c]{\textwidth}
\textbf{Problem 1:} Approximate a matrix element in the Young-Yamanouchi
basis of an irreducible representation of the symmetric group $S_n$. \\
\textbf{Input:} A Young diagram $\gamma$ of two rows specifying the
irreducible representation, a permutation from $S_n$, a pair of
standard Young tableaux of shape $\gamma$ indicating the desired
matrix element, and a polynomially small parameter $\epsilon$.\\
\textbf{Output:} The specified matrix element to within $\pm
\epsilon$. \\
\end{minipage}
To do this we simply translate the pair of Young tableaux into a pair
of $j$-labeled binary trees $\lambda$ and $\lambda'$, as described in figure
\ref{correspondence}. Then we build the tree states $\ket{\lambda}$,
permute the particles according to $\pi$, and obtain the approximate
$\ket{\lambda'}$-amplitude of the resulting state using the Hadamard test.

No polynomial time classical algorithm is known for problem 1. There
is a small body of literature on optimized classical algorithms for
computing matrices from Young's orthogonal form\cite{Hamermesh,
Boerner, Wu_Zhang1, Wu_Zhang2, Clifton, Rettrup, Pauncz},
mainly for applications to computational chemistry. All of these
algorithms have worst-case runtime that scales exponentially in
$n$. To be fair, it should be noted that these classical algorithms
are numerically exact, and compute all of the matrix elements at
once. It is not clear how much effort has gone into fast classical
algorithms for approximating individual matrix elements from Young's
orthogonal form. Nevertheless, it seems likely that a polynomial time
algorithm for problem 1 presents a genuine exponential speedup over
classical computation. A thorough discussion of this point is given in
\cite{Jordan_groups}.

Lastly, we note that problem 1 is a complete problem for the
subclass of permutational quantum computations in which the state
preparation and measurement are both fixed to be trees of the type
shown in (\ref{centipedes}). In the next two sections we describe a different
problem, which is complete for the reverse situation: the permutation
is fixed to be the identity, and the initial and final trees are free
to be chosen arbitrarily. 

\section{The Ponzano-Regge Model}

The Ponzano-Regge model\cite{PR} is a 3-dimensional topological
quantum field theory (TQFT) for a class of triangulated manifolds. It is often
studied in the context of quantum gravity (see appendix
\ref{gravity}), but it is a mathematical object of intrinsic interest,
and in particular it gives rise to a nontrivial three-manifold
invariant. In this section we describe the Ponzano-Regge model, and to
do so we first give a brief overview of topological quantum field
theories in general.

The term ``topological quantum field theory'' is used in the
literature to refer to two related but distinct concepts. It is first
of all used to refer, somewhat loosely, to any quantum field theory in
which the action is diffeomorphism invariant. Perhaps the best known
example is Chern-Simons theory. A second, more mathematical definition
of the term is any structure satisfying the Atiyah axioms,
proposed in\cite{Atiyah}. We use the second definition throughout this
paper. The two concepts are not unrelated. The matrix elements of
the linear transformation corresponding to a cobordism (described
below) are analogous to the transition amplitudes that one would
compute using a path integral in more conventional formulations of
quantum field theory. For a complete and mathematically precise
description of axiomatic TQFT see \cite{Atiyah}.

Essentially, an $n$-dimensional axiomatic topological quantum field
theory (TQFT) is a map that associates a Hilbert space to any $(n-1)$-manifold,
and to any $n$-dimensional manifold ``interpolating'' between a pair
of $(n-1)$-dimensional manifolds, it associates a linear transformation
between the corresponding Hilbert spaces. More precisely, a cobordism
is defined to be a triple $(M,A,B)$ where $M$ is an $n$-manifold whose
boundary is the disjoint union of $(n-1)$-manifolds $A$ and $B$. This
provides a well-defined notion of an ``interpolation'' between $A$ and
$B$. For example, the circle $S^1$ is a 1-manifold, and a tube $S^1
\times [0,1]$ is a cobordism between two circles. A 2-dimensional TQFT
associates a Hilbert space $\mathcal{H}_{S^1}$ to $S^1$ and a linear
transformation $M_0:\mathcal{H}_{S^1} \to \mathcal{H}_{S^1}$ to the tube. 
\[
\includegraphics[width=0.24\textwidth]{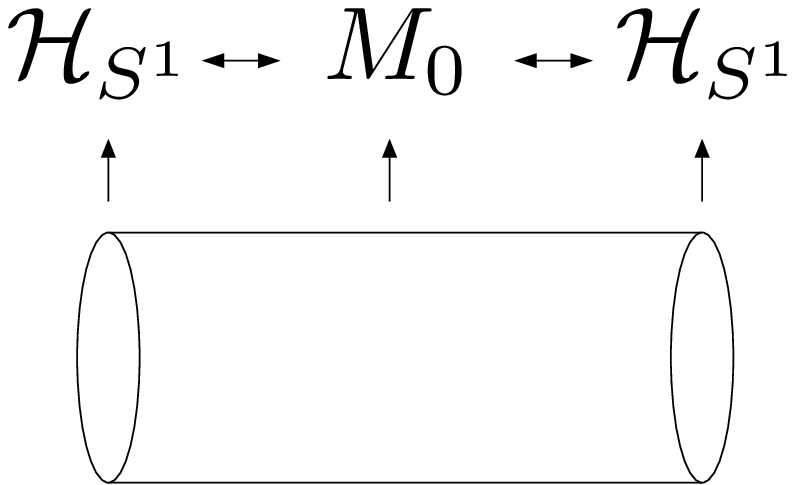}
\]
A different cobordism between the same pair of boundaries may be mapped to a
different linear transformation between the same pair of Hilbert spaces.
\[
\includegraphics[width=0.24\textwidth]{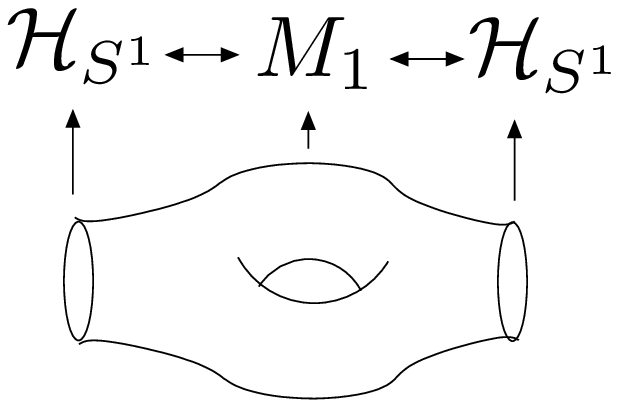}
\]
If we compose together two cobordisms, we compose the corresponding
linear transformations.
\begin{equation}
\label{functor}
\includegraphics[width=0.36\textwidth]{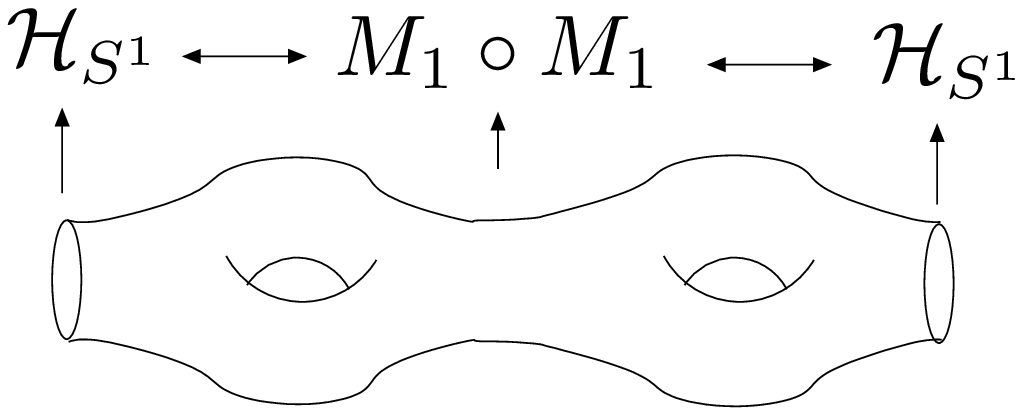}
\end{equation}
Mathematicians express this property by saying that a TQFT is a
``functor''. The linear transformation associated to a cobordism by a
TQFT depends only on the topology of the cobordism, not the geometric
details. Therefore we can see, for example, that $M_0 \circ M_0 =
M_0$. To the empty boundary $\emptyset$ we associate the Hilbert space
$\mathbb{C}$. We can think of a closed manifold as a cobordism between
$\emptyset$ and $\emptyset$. Therefore an $n$-dimensional TQFT
associates to any closed $n$-manifold a map from $\mathbb{C}$ to
$\mathbb{C}$, that is, a complex number. This map is a
$\mathbb{C}$-valued topological invariant of closed $n$-manifolds.

More precisely, let $f$ be a function from the set of
$n$-manifolds $M_n$ to some other set $S$. If $f$ has the property
that $f(A) = f(B)$ whenever $A,B \in M_n$ are equivalent
(homeomorphic) then $f$ is a $n$-manifold invariant. Note that the
definition does not require $f(A) \neq f(B)$ whenever $A$ is
nonhomeomorphic to $B$. An $n$-manifold invariant with this latter
property is said to be complete.

The Ponzano-Regge model associates linear transformations to
3-manifolds, which can be thought of as cobordisms between
2-manifolds. There are several ways of describing three 
manifolds\footnote{The interested reader should
  look up Heegaard splittings and surgery presentations.}, but perhaps
the most intuitive is by triangulation. A triangulated
3-manifold is a list of tetrahedra and a list of which face is ``glued''
to which. For example, we could take two tetrahedra and glue their
faces as follows.
\[
\includegraphics[width=0.37\textwidth]{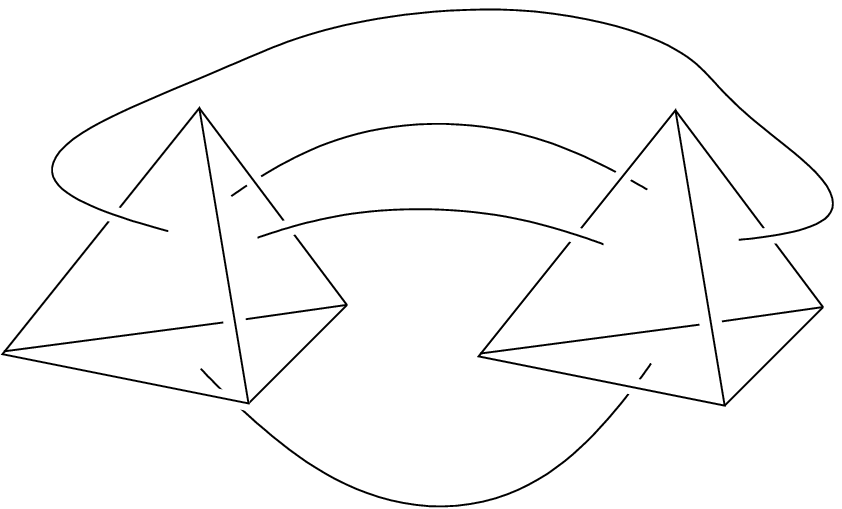}
\]
The term triangulation arises from 2-manifolds, which can be specified
by a list of triangles and a list of which edge is glued to which
edge. It is common to use ``triangulation'' to describe the analogous
concept in any dimension even though, for example, triangulated
three-manifolds are made from tetrahedra rather than triangles. 

A given 3-manifold can be triangulated in many different ways. Because
we only care about the topology of the manifold, finer triangulations
are in no way preferable to coarser ones. The question of which
triangulations specify the same (homeomorphic) manifolds is completely
answered by Pachner's theorem.

\begin{figure}
\begin{center}
\includegraphics[width=0.35\textwidth]{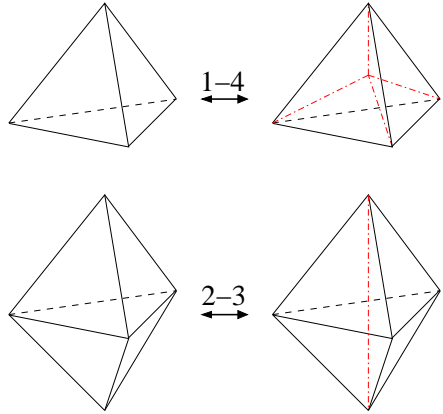}
\caption{\label{moves} The 1-4 move takes a tetrahedron and subdivides
it into four tetrahedra by introducing a new vertex in the center. The
2-3 move take two tetrahedra sharing a face, removes the shared face,
and slices the resulting octahedron longitudinally into three tetrahedra.}
\end{center}
\end{figure}

\begin{theorem}[Pachner's theorem\cite{Pachner}] Two triangulations
  specify the same 3-manifold if and only if they are connected by a
  finite sequence of the 2-3 and 1-4 moves and their inverses, as
  illustrated in figure \ref{moves}.
\end{theorem}

Although conceptually and mathematically useful, Pachner's theorem
does not directly yield an algorithm for the problem of deciding
equivalence of triangulated manifolds. One can search for a sequence
of Pachner moves connecting a pair of triangulations, but one
does not know when to give up the search, because no upper bound is
known on the number of necessary moves. Eventually it was
proven\footnote{According to \cite{Gowers}, the proof is highly
  nontrivial and makes use of tools developed by Perelman
  in his proof of the Poincar\'e conjecture.} that 
the equivalence problem for 3-manifolds is algorithmically decidable
(see \cite{Gowers}), but the problem 
is not known to be in P. The equivalence problem for $n$-manifolds for
any $n \geq 4$ is undecidable\cite{Markov}. The equivalence problem
for orientable 2-manifolds is
in P (see appendix \ref{twoman}). Lacking an efficient way to decide
equivalence of three-manifolds it is natural to look for partial
solutions such as those provided by 3-manifold invariants. 

In the Ponzano-Regge model, we are given a triangulated 3-manifold. We
then associate one $j$-variable to each edge of each
tetrahedron. These $j$-variables represent spins and take 
integer and half-integer values. To a closed manifold the
Ponzano-Regge model associates the following amplitude\cite{Barrett}.
\begin{equation}
\label{Wclosed}
Z_{\mathrm{closed}} = \sum_{j-\mathrm{labelings}}
\prod_{\mathrm{edges}} (-1)^{2j}
(2j+1) \prod_{\mathrm{faces}} (-1)^{j_1+j_2+j_3}
\prod_{\mathrm{tetrahedra}} \left\{ \begin{array}{ccc}
j_1 & j_2 & j_3 \\
j_4 & j_5 & j_6 \end{array} \right\}
\end{equation}
The value of the $6j$-symbol is not invariant under the $6!$
permutations of the indices. To get the correct Ponzano-Regge
amplitude we must associate the indices in the $6j$ symbol to the
edges of a tetrahedron as follows.
\begin{equation}
\label{ordering}
\includegraphics[width=0.4\textwidth]{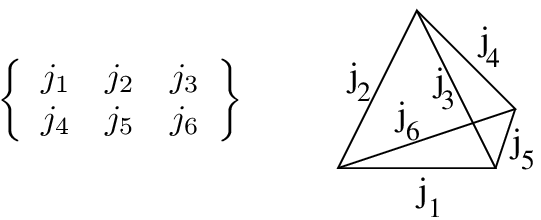}
\end{equation}
The $6j$ symbol is invariant under the 24 symmetries of a tetrahedron,
which shows that this procedure is consistent. The diagram above just
indicates that each pair of indices sharing a column in the $6j$
symbol must correspond to a pair of nonadjacent (\emph{i.e.} opposite)
edges in the tetrahedron.

In principle, there are infinitely many $j$-labelings to sum over in
equation \ref{Wclosed}. However, the $6j$-symbol is only nonzero if
the triple of $j$-values associated to each face is admissible by the
rules of quantum angular momentum addition, described in equation
\ref{laws}. Thus, in some cases, the Ponzano-Regge amplitude contains
only finitely many nonzero terms. In the case of infinitely many
admissible labelings, the sum often diverges. We return to the
question of which triangulations yield finite amplitudes below, but
first we describe the Ponzano-Regge amplitude associated to a
3-manifold with boundary.

Let $M$ be a 3-manifold whose boundary is the union of the
2-manifolds $A$ and $B$. That is, $M$ is a generalized\footnote{In a
  true cobordism, the boundary of $M$ is the \emph{disjoint} union of
  $A$ and $B$.} cobordism between $A$ and $B$. The triangulation (by tetrahedra) of
$M$ induces triangulations (by triangles) of $A$ and $B$. Each edge of
a triangulation of $M$ is either internal, on the boundary $A$, on
the boundary $B$, or on both $A$ and $B$. The same applies to the
faces. To any labeling of the boundaries, the Ponzano-Regge model
associates the following amplitude, where the sum is over
$j$-labelings of internal edges.
\begin{eqnarray*}
Z & = & \sum_{j-\mathrm{labelings}} \left( \prod_{e \in \mathrm{edges}} \left[
  (-1)^{2j_e} (2j_e+1) \right]^{(2-b(e))/2} \times \right.\\
 & & \left. \prod_{f \in \mathrm{faces}} \left[
  (-1)^{j_{f(1)}+j_{f(2)}+j_{f(3)}} \right]^{(2-b(f))/2}
\prod_{t \in \mathrm{tetrahedra}} \left\{ \begin{array}{ccc}
j_{t(1)} & j_{t(2)} & j_{t(3)} \\
j_{t(4)} & j_{t(5)} & j_{t(6)} \end{array} \right\} \right)
\end{eqnarray*}
Here $b(e)$ is the number of boundaries on which the edge $e$
lies, 0, 1, or 2. Similarly, $b(f) \in \{0,1,2\}$ is the number of
boundaries on which the face $f$ lies. $\{f(1),f(2),f(3)\}$ are the
three edges of face $f$, and $\{t(1),t(2),t(3),t(4),t(5),t(6)\}$ are
the six edges of tetrahedron $t$. We recover equation \ref{Wclosed} as
the special case when all edges and faces are internal ($b(e) = 0 \
\forall e$ and $b(f) = 0 \ \forall f$).

To each boundary we can associate a (possibly infinite-dimensional)
Hilbert space corresponding to all admissible labelings of the
triangulation. The Ponzano-Regge amplitudes between a pair of
labelings is thus a matrix element of a linear transformation between
these Hilbert spaces. This mapping from cobordisms to linear
transformations obeys the functoriality property \ref{functor}. That
is, when we compose two cobordisms, we compose the corresponding
linear transformations. To see this, note that by the usual rule of
matrix multiplication, we
sum over the labelings of the edges that we have glued. This is
consistent with the prescription for calculating the Ponzano-Regge
amplitude for the resulting single cobordism because these edges
have now become internal. Furthermore, each edge $e$ has one factor of
$\sqrt{(-1)^{2j_e}(2j_e+1)}$ from each of the two matrices we have
multiplied, resulting in a factor of $(-1)^{2j_e}(2j_e+1)$, as befits
an internal edge. Similarly, the faces that are glued pick up one
factor of $\sqrt{(-1)^{j_{f(1)}+j_{f(2)}+j_{f(3)}}}$ from each of the
linear transformations being composed, giving them the correct
weighting for an internal face.

For a three-dimensional quantum field theory on triangulated manifolds
to be topological, it should be independent of triangulation, that is,
invariant under the Pachner moves. The Ponzano-Regge model fails to be
topological in general. It is invariant under the 2-3 Pachner move as
a consequence of the Beidenharn-Elliot identity for $6j$
symbols\cite{mathworld}:
\[
\sum_x (-1)^\phi (2x+1) 
\left\{ \begin{array}{ccc} a & b & x \\ c & d & g \end{array} \right\}
\left\{ \begin{array}{ccc} c & d & x \\ e & f & h \end{array} \right\}
\left\{ \begin{array}{ccc} e & f & x \\ b & a & j \end{array} \right\}=
\left\{ \begin{array}{ccc} j & h & j \\ e & a & d \end{array} \right\}
\left\{ \begin{array}{ccc} g & h & j \\ f & b & c \end{array} \right\},
\]
where $\phi = a+b+c+d+e+f+g+h+x+j$. However, it is not invariant under
the 1-4 move. In fact, by applying the 1-4
move one can go from a triangulation whose Ponzano-Regge amplitude has
finitely many terms to one whose amplitude has infinitely many
terms. A Ponzano-Regge amplitude with infinitely many terms could
still be convergent, but generically this seems not to be the
case\cite{Barrett}.

In a Ponzano-Regge summation for a 3-manifold with boundary, the
boundary conditions (\emph{i.e.} the $j$-labels on the boundary edges)
limit the admissible labels on the internal edges. For some
triangulations this limitation ensures that there are only finitely
many possible labels on every edge. For other triangulations there
exist some internal edges whose label is not constrained to a finite
set by the boundary conditions. Barrett and Naish-Guzman call the set
of edges whose labels are not constrained to a finite set the
``tardis'' of the triangulation. The tardis depends only on
the manifold and its triangulation, not on the particular $j$-values
assigned to the boundary. The Ponzano-Regge amplitudes
between boundary labellings of a triangulation whose tardis is
the empty set (called a non-tardis triangulation in \cite{Barrett}) are
finite. Furthermore, as proven in \cite{Barrett},
\begin{theorem}{\cite{Barrett}}
Let $M$ be a manifold with a fixed triangulation of its boundary. Then
any two non-tardis triangulations compatible with the given boundary
triangulation yield the same Ponzano-Regge amplitudes.
\end{theorem}
For the class of non-tardis triangulations the Ponzano-Regge model
thus acts as a topological quantum field theory. It is also worth
noting that given a triangulation with $n$ tetrahedra, a classical
computer can decide whether it is non-tardis in $\mathrm{poly}(n)$
time\cite{Barrett}.

\section{Approximating Ponzano-Regge Transition Amplitudes}
\label{PR_algorithm}

In this section we show that for a certain subclass of non-tardis
triangulations the Ponzano-Regge transition amplitudes can be
efficiently approximated to polynomial additive precision on a
permutational quantum computer. 

A single tetrahedron is a triangulation of the 3-ball. Its boundary is
a triangulated 2-sphere. We can divide the 2-sphere boundary into a
pair of discs, and think of the 3-ball as a generalized cobordism
between these two discs. We can do this such that each disc is
triangulated into two triangles, as shown below.
\begin{equation}
\label{tet}
\includegraphics[width=0.1\textwidth]{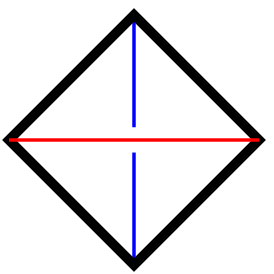}
\end{equation}
The six lines in diagram \ref{tet} are the edges of a tetrahedron. The
vertical line lies in the top disc, the horizontal line lies in the
bottom disc, and the thick lines around the perimeter lie on both
discs. The Ponzano-Regge model associates an infinite-dimensional
Hilbert space $\includegraphics[width=1.5em]{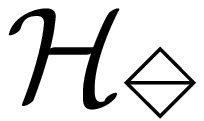}$ to the
two-face triangulation of a disc, namely the formal span of all
admissible labelings of the five edges of the
triangulation. Corresponding to the tetrahedron we have a linear 
transformation $M_{\mathrm{tet}}:
\includegraphics[width=1.5em]{hdiamond.eps} \to
\includegraphics[width=1.5em]{hdiamond.eps}$. $M_{\mathrm{tet}}$ is
completely specified by the matrix elements between each pair of
admissible labellings of the disc triangulations. By the definitions
given in the previous section, these matrix elements are
\begin{eqnarray}
\label{fullform}
\Bra{\begin{array}{c} \includegraphics[width=10mm]{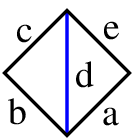}
  \end{array}} M_{\mathrm{tet}}
\Ket{\begin{array}{c} \includegraphics[width=10mm]{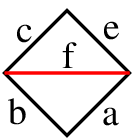}
  \end{array}} & = & \sqrt{(-1)^{2f} (2f+1)} \sqrt{(-1)^{2d} (2d+1)}
\times \\ \nonumber
& & \sqrt{(-1)^{c+b+d}} \sqrt{(-1)^{d+e+a}} \sqrt{(-1)^{c+e+f}}
\sqrt{(-1)^{b+a+f}} \left\{ \begin{array}{ccc} a & b & f\\
c & e & d \end{array} \right\}
\end{eqnarray}
Simplifying this expression, we recognize it as
\begin{equation}
\label{simplified}
\left[ \begin{array}{ccc} a & b & f \\ c & e & d \end{array}
\right],
\end{equation}
the recoupling tensor from equation \ref{recoupling_tensor}. This
observation generalizes as follows.

\begin{figure}
\begin{center}
\includegraphics[width=0.7\textwidth]{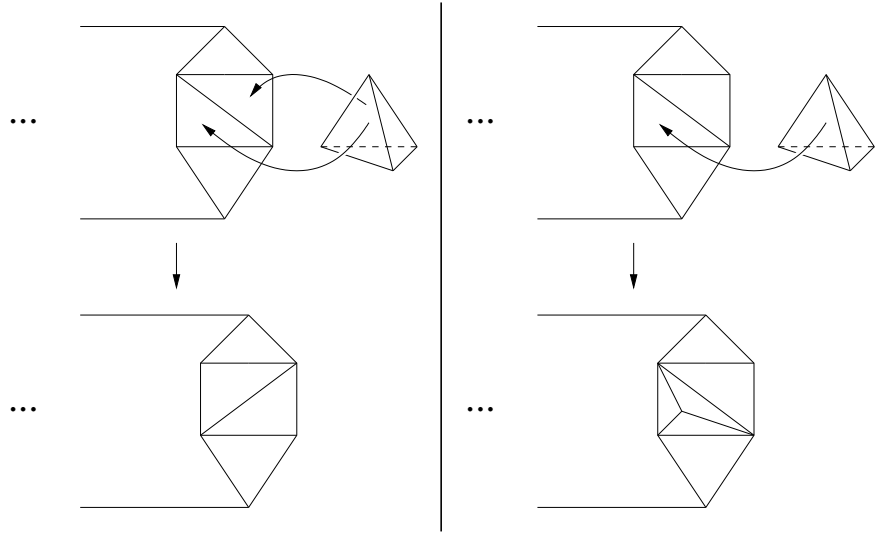}
\caption{\label{retriangulate} On the left we glue two adjacent
  faces of a tetrahedron to two adjacent triangles of the boundary of
  a 3-manifold. The result is a new triangulation of the boundary in
  which one edge has been flipped. On the right we instead glue only
  one face of the tetrahedron. In the resulting triangulation one of
  the triangles gets split into three.}
\end{center}
\end{figure}

\begin{figure}
\begin{center}
\includegraphics[width=0.45\textwidth]{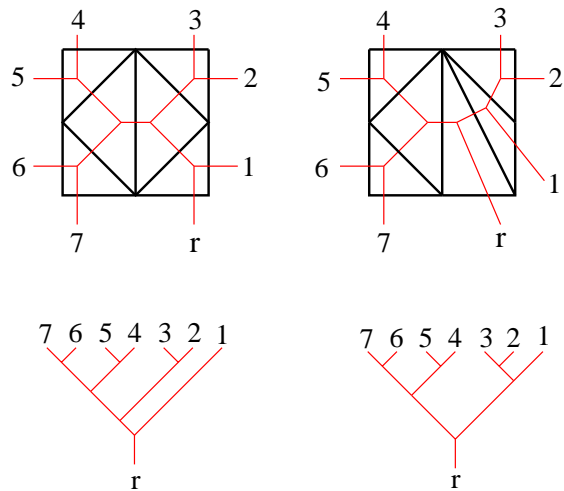}
\caption{\label{dualtreemove} Flipping an edge of a triangulation
induces a change on the corresponding dual triangulation. If the dual
triangulation is a tree, then this corresponds to a recoupling of
spins. Here we have chosen one edge as the root and labeled it
``r''. We then draw the dual tree in more conventional shape to
illustrate the connection to recoupling moves discussed in section
\ref{model}. We have added arbitrary labels to the other leaves for
clarity.}
\end{center}
\end{figure}

We can build up a 3-manifold by gluing on additional tetrahedra one by
one. As we do so, the triangulation of the surface gets modified. The
change in triangulation depends on how many faces of the new
tetrahedron we glue to existing faces, as illustrated in figure
\ref{retriangulate}. Thus the gluing of a tetrahedron can be thought
of as a cobordism or as a retriangulation of the boundary of a
three-manifold. In particular if we glue two faces of a tetrahedron
we correspondingly flip one edge of the triangulation.

The dual to a $j$-labeled triangulation is another $j$-labeled graph,
as illustrated in figure \ref{dualtreemove}. If this $j$-labeled dual
graph is a tree, then the flip move on the original graph translates
to a recoupling move on the dual tree. (See figure
\ref{dualtreemove}.) The linear transformation that the Ponzano-Regge
model associates to this cobordism is identical to the recoupling
transformation of the corresponding $j$-labeled dual tree. 

In the permutational model we can efficiently find a polynomial
additive approximation to the matrix elements between any pair of
$j$-labeled binary trees in time polynomial in the number of
leaves. Thus, the following problem can be solved on a permutational
quantum computer in $\mathrm{poly}(n,m,1/\epsilon)$ time.\\ \\
\noindent
\begin{minipage}[c]{\textwidth}
\textbf{Problem 2:} Approximate Ponzano-Regge transition amplitude. \\
\textbf{Input:} We are given two triangulated surfaces $A$ and $B$
such that the dual to the triangulations are both binary trees of $n$
leaves. We are also given sequence of $m$ pairs of neighboring triangles
on which to glue two faces of successive tetrahedra. This induces a
sequence of edge flips taking us from triangulation $A$ to
triangulation $B$. We are given $j$-labelings for $A$ and $B$. Lastly, 
we are given positive parameter $\epsilon$.\\
\textbf{Output:} The real and imaginary parts of Ponzano-Regge
transition amplitude corresponding to the above cobordism, to within
$\pm \epsilon$. \\
\end{minipage}

Note that the class of cobordisms described in problem 2 is quite
restricted. In general triangulated surfaces may not have duals
that are binary trees. Furthermore, as illustrated in figure
\ref{retriangulate}, there are ways to glue tetrahedra onto a surface
other than two faces at a time. It is an open problem whether
permutational quantum computers can efficiently approximate
Ponzano-Regge transition amplitudes for a more general class of
cobordisms. On the other hand, it is clear that problem 2 is a
complete problem for the ``pure recoupling'' class of permutational
quantum computations, in which the binary trees are chosen
arbitrarily, but the permutation is fixed to be the identity.

\section{Mixed States}
\label{mixed}

In this section we consider a modified version of permutational quantum
computation in which the initial state is highly mixed. We then show
that the resulting complexity class is contained in BPP. In contrast,
if we take the standard quantum circuit model and analogously apply it
to highly mixed initial states, we obtain a complexity class DQC1
which appears to extend beyond BPP, although it is probably weaker
than BQP. We start by describing DQC1.

In the standard quantum circuit model, one begins with a canonical
pure state, such as $\ket{0}^{\otimes n}$, applies a quantum circuit,
and then performs a simple measurement, such as measuring each qubit
independently in the $\{ \ket{0},\ket{1} \}$ basis. Experimentally,
it is often difficult to obtain the initial pure state
$\ket{0}^{\otimes n}$. In particular, several of the early quantum
computing experiments used liquid state NMR, in which the quantum
states being manipulated are highly mixed. This led to some debate
as to whether the NMR experiments were truly quantum computation.

To address this issue, Knill and Laflamme introduced an idealized
model of quantum computation on highly mixed states, called the one
clean qubit model\cite{Knill_DQC1}. In this model one is given the
initial state $\rho$ in which one qubit is in a pure state and the
remaining $n$ qubits are maximally mixed.
\[
\rho = \ket{0} \bra{0} \otimes \frac{I}{2^n}
\]
Then one is allowed to apply polynomially many quantum gates to this
state, and lastly measure the first qubit in the $\{ \ket{0},\ket{1} \}$
basis. The complexity class of problems solvable in polynomial time in
this model is called DQC1.

\begin{figure}
\begin{center}
\includegraphics[width=0.25\textwidth]{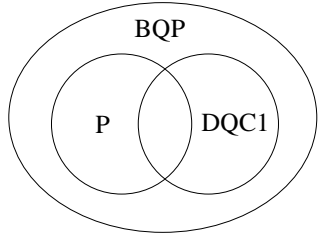}
\caption{\label{mixed_complexity} This diagram shows the conjectured
  relationships between classical polynomial time (P), quantum
  polynomial time (BQP), and one clean qubit (DQC1). P and DQC1 are
  known rigorously to be contained in BQP but the containments are
  not known rigorously to be strict.} 
\end{center}
\end{figure}

Let $C$ be a quantum circuit on $n$ qubits and let $U_C$ be the
corresponding $2^n \times 2^n$ unitary matrix. The problem of
estimating $\tr(U_C)$ is DQC1-complete\cite{Knill_DQC1,
  Shepherd}. A certain problem of estimating Jones polynomials is also
DQC1-complete\cite{Shor_Jordan}, and the generalization to HOMFLY
polynomials is also efficiently solvable on one clean qubit
computers\cite{Jordan_Wocjan}. No polynomial time algorithms for 
these problems are known. On the other hand, most evidence
suggests that one clean qubit computers are less powerful than
standard quantum computers\cite{Ambainis_DQC1, Datta}. Just as with
PQP, one must be careful in choosing what sort of computer controls
the experiments performed on the qubits. If this computer is a P
machine then DQC1 automatically contains P. To make a more meaningful
comparison to classical computation it is better to choose a weaker
computational model such as L or NC1 to control the experiments (see
\cite{complexity_zoo} for the definitions of these complexity
classes). The conjectured relationships between P, BQP, and DQC1 are
illustrated in figure \ref{mixed_complexity}.

The one clean qubit model is not necessarily physically realistic in
its details. Nevertheless, it provides a proof of principle that
quantum computation on highly mixed states is still probably capable
of achieving exponential speedups over classical computers for certain
problems. Furthermore, it provides perhaps the only model of quantum
computation other than PQP which yields a complexity class apparently
distinct from BQP but not contained in P. 

We can similarly formulate a one clean qubit version of the
permutational model. By analogy to DQC1 we define the one clean
qubit version of PQP to be the set of problems efficiently solvable
given the ability to approximate the trace of irreducible
representations of $S_n$ (\emph{i.e.} characters) to $\pm \epsilon$
precision in $\mathrm{poly}(n,1/\epsilon)$ time. Such a definition
seems mathematically natural. Furthermore, it is equivalent to a
somewhat natural physical model, as follows. We start with the initial
state
\[
\rho_J = \ket{\psi} \bra{\psi} \otimes \frac{1}{d_J} \sum_{x=1}^{d_J}
\ket{J,x}\bra{J,x}
\]
on $n+1$ spin-1/2 particles. Here $\ket{\psi} = \alpha \ket{0} + \beta
\ket{1}$ is an arbitrary pure state on one spin, and
$\ket{J,1},\ket{J,2},\ldots,\ket{J,d_J}$ is a complete basis for the
space of states of $n$ spins with total angular momentum $J$. Thus
these $n$ spins are in the maximally mixed state within the 
subspace of total angular momentum $J$. We then perform a coherently
controlled permutation on the $n$ maximally mixed spins, and measure
the back-action on the first spin, as illustrated below. 
\[
\mbox{\Qcircuit @C=1em @R=.7em {
  \lstick{\alpha \ket{0}+\beta \ket{1}} & \qw & \ctrl{1} & \gate{H} & \meter \\
  \lstick{\frac{1}{d_J} \sum_{x=1}^{d_J} \ket{J,x}\bra{J,x}} & {/} \qw  &
  \gate{U_\pi} & {/} \qw & \qw & }}
\]
This type of interferometric measurement is standard in quantum
computing and is known as the Hadamard test. The final measurement is
in the $\{\ket{0}, \ket{1}\}$ basis. The probability of outcome
$\ket{0}$ is
\[
p_0 = \frac{1}{2} \left[ 1 + 2 \mathrm{Re} \left( \frac{ \alpha^* \beta
      \sum_{x=1}^{d_J} \bra{J,x} U_\pi \ket{J,x}}{d_J} \right) \right].
\]
As discussed in section \ref{model}, $U_\pi$ acts on the span of
$\ket{J,1},\ket{J,2},\ldots,\ket{J,d_J}$ as an irreducible
representation of $S_n$. Specifically, $U_\pi$ acts as the irreducible
representation corresponding to a Young diagram of two rows where the 
overhang of the top row over the bottom row is $2J$. Thus,
\[
p_0 = \frac{1}{2} \left[ 1 + 2 \mathrm{Re} \left( \alpha^* \beta
    \frac{\chi_J}{d_J} \right) \right],
\]
where $\chi_J$ is the character of this irreducible representation. 

By performing the Hadamard test $O(1/\epsilon^2)$ times with
$\alpha=\beta=\frac{1}{\sqrt{2}}$ one can estimate $\mathrm{Re} \left(
  \frac{\chi_J}{d_J} \right)$ to within $\pm \epsilon$. Similarly,
choosing $\alpha = \frac{1}{\sqrt{2}}$, $\beta = - \frac{i}{\sqrt{2}}$
yields $\mathrm{Im} \left( \frac{\chi_J}{d_J} \right)$. Thus this
model efficiently solves the problem of estimating
$\frac{\chi_J}{d_J}$ to polynomial additive precision. Furthermore, simulating
this model reduces to the problem of estimating $\frac{\chi_J}{d_J}$
to polynomial additive precision. Thus this problem is complete for
the one clean qubit version of permutational quantum computation. Note
that the hardness result holds for any choice of projective
measurement, not just
$\begin{array}{c}\mbox{\Qcircuit @C=.8em @R=.6em { & \gate{H} & \meter
  }} \end{array}$.

As shown in \cite{Jordan_groups}, the normalized character of any
irreducible representation of the symmetric group can be approximated
to $\pm \epsilon$ with probability $1-\delta$ in
$\mathrm{poly}(n,1/\epsilon,\log(1/\delta))$ time on a classical
computer using an algorithm based on random sampling. This shows that
the one clean qubit version of PQP is contained in the complexity
class BPP. This contrasts with DQC1, which is unlikely to be contained
in BPP. One could interpret this as a form of indirect evidence that
PQP is weaker than BQP.

An arguably more compelling comparison can be made to topological
computation. The problem of estimating a particular matrix element of
the Fibonacci representation of the braid group $B_n$ is
BQP-complete\cite{Freedman,AJL,Aharonov_Arad}. Estimating the characters of the
Fibonacci representation is DQC1-complete\cite{Shor_Jordan}. The Fibonacci
representation of the braid group is closely related to Young's
orthogonal representation of the symmetric group. More precisely, as
discussed in \cite{Kauffman}, the former is a $q$-deformation of the
latter. The fact that estimating characters of $S_n$ is
easier than estimating the characters of the corresponding Fibonacci
representation of $B_n$ suggests that estimating matrix
elements of Young's orthogonal form may be easier than estimating matrix
elements of the Fibonacci representation. That is, it suggests that
estimating matrix elements of Young's orthogonal form may not be
BQP-complete.

\section{Fault Tolerance}
\label{impervious}

To implement a permutational quantum computation, one prepares spins
into states of known total angular momentum, moves the spins around,
and measures total angular momentum. For a fixed number of spins, the
set of allowed operations is discrete and finite. Errors in the
trajectories of the spins do not cause computational errors unless
the trajectory is so deformed as to implement a different permutation
than that intended. In this respect, the permutational model is
similar to the topological model. In the topological model, trajectory
errors do not cause computational errors unless they are large enough
to induce a braid other than the one intended.

There is a second, more subtle way in which permutational quantum
computers are resistant to error; they are impervious to the effects
of uniform magnetic fields. A uniform magnetic field $\vec{B}$ acts on $n$
spin-1/2 particles through the Hamiltonian
\[
H_{\vec{B}} = \vec{B} \cdot \sum_{i=1}^n \vec{S}_i,
\]
where $\vec{S}_i$ is the angular momentum operator on spin $i$, as
described in equation \ref{vecS}. Applying this Hamiltonian for time
$t$ induces the unitary transformation
\[
U_{\vec{B}}(t) = e^{-i H_{\vec{B}} t} = u^{\otimes n}
\]
for some $u \in SU(2)$.

The unitary transformation $u^{\otimes n}$ commutes with the total
angular momentum operator $S_b$ for any subset $b$ of the $n$ spin-1/2
particles. Thus, it does not affect the computation. Recall from
section \ref{model} that in addition to total angular momenta of various
subsets of spins, we need the operator
\[
Z_{\mathrm{total}} = \frac{1}{2} \sum_{i=1}^n \sigma_z^{(i)}
\]
to obtain a complete set of commuting observables. $u^{\otimes n}$
affects only the degree of freedom described by the eigenvalue of this
operator. (This is an example of Schur-Weyl duality\cite{Bacon}.) In
the language of quantum error correction, the space of $j$-labeled
tree states is a noiseless subsystem\cite{Lidar_Whaley} with respect to
errors of the form $u^{\otimes n}$.

\section{Concluding Remarks}
\label{conclusion}

In this paper we analyze a model of quantum computation
based on the permutation of spin-1/2 particles. We call the set of
problems solvable in this model in polynomial time
PQP. Permutational computers can be simulated efficiently by standard
quantum computers. Thus PQP is contained in BQP. On the other hand, we
have presented two quantum algorithms for permutational quantum
computers that seem to exhibit exponential speedup over classical
computation. Thus it seems unlikely that PQP is contained in P.

The question of whether PQP is equal to BQP remains open. In other
words, we do not know whether permutational quantum computers are
universal. To begin to address this question, we first review the
techniques that were used to prove the universality of the quantum
circuits and topological quantum computers.  

In principle, the set of unitary transformations that one might wish to
perform on $n$ qubits is a compact but continuous group
$SU(2^n)$. In contrast, the set of quantum circuits achievable with a
finite gate set is discrete. Nevertheless, certain finite sets of
gates achieve universality in the sense that the discrete infinite set
of unitaries achievable by composing these gates into quantum circuits
is dense in $SU(2^n)$ \cite{Nielsen_Chuang}. Thus any desired unitary
in $SU(2^n)$ can be approximated to any desired level of precision by
a sufficiently large quantum circuit. To approximate an arbitrary
unitary on $n$ qubits to $1/\mathrm{poly}(n)$ precision requires
exponentially many gates in general. Most research on quantum
computation focuses on the small subset of $SU(2^n)$ implementable by
circuits of $\mathrm{poly}(n)$ gates. Using the Solovay-Kitaev
theorem\cite{Nielsen_Chuang}, one can show that this subset does not
depend on the choice of gate set. That is, any universal gate set can
simulate any other with only logarithmic overhead.

Similarly, the braid group $B_m$ on $m$ strands is a discrete infinite
group for any fixed $m$. As shown in \cite{Freedman, Freedman2, AJL,
  Aharonov_Arad}, the unitary representations of $B_m$ induced by
certain anyons are dense in the corresponding unitary
groups. Furthermore, these authors show that the set of unitaries
approximable by braids of polynomially many crossings coincides with
the set of unitaries approximable by quantum circuits of polynomially
many gates.

The permutational model of quantum computation is very different from
the quantum circuit model and anyonic models in that the set of
unitaries achievable with $n$ spins is finite. As discussed in
\cite{Marzuoli}, the number coupling schemes for $n$ spins, which we
represent as binary trees of $n$ leaves, is $C_{n-1}$, where $C_n$ is
the $n\th$ Catalan number 
\[
C_n = \frac{(2n)!}{(n+1)!n!}.
\]
When specifying a permutational computation we choose two such
couplings and one of the $n!$ possible permutations. Thus, the number
of implementable unitary transformations is upper bounded by
$n!C_{n-1}^2$. By Stirling's approximation this is $2^{O(n \log
  n)}$. In contrast, suppose we want to choose a finite subset $S$ of
$SU(d)$ such that for any $u \in SU(d)$ there exists $v \in S$ such
that $\|v-u\| < \epsilon$. As discussed in \cite{mythesis}, for fixed
$\epsilon$ the smallest set $S$ satisfying this density condition has
exponentially many elements as a function of $d$. The unitary
transformations of an $n$-spin permutational computer act on a Hilbert
space whose dimension is exponential in $n$. Any dense subset of the
unitaries on this Hilbert space would therefore have a doubly
exponential number of elements. Thus the set of $2^{O(n \log n)}$
unitaries actually implementable on a permutational quantum computer
is very far from dense.

The limited number of unitaries implementable by a permutational
quantum computer with a fixed number of spins means that the standard
universality arguments based on density cannot work. If permutational
computers are universal their method of simulating quantum circuits
will have to use a number of spins that scales not only with the
number of qubits on which the quantum circuit acts, but also on the
number of gates in the circuit. Arguably the failure of standard
techniques of proving universality and the fact that the one clean
qubit version of permutational quantum computation is weaker than the
one clean qubit version of standard quantum computation suggest
that PQP is in fact weaker than BQP.

The problem of whether PQP even contains all of P is also an
open question. It could be that PQP and P are incomparable, that is,
classical computers and permutational quantum computers can each solve
some problems in polynomial time that the other cannot. Note that, as
discussed in section \ref{model}, analysis of this question requires a
careful formulation of PQP.

The physical implementation of permutational quantum computers raises many
additional questions. For one, the problem of finding a
physically realistic method to implement the interferometric
measurements discussed in section \ref{model} remains
unsolved. Without these we obtain a slightly weaker version of
permutational quantum computation in which the phase information is
unrecoverable.

In addition, one could investigate potential implementations of the
permutational model other than the actual manipulation of spin-1/2
particles. One possibility is to find a material whose quasiparticles
obey the ``braiding'' and recoupling rules described in \ref{recouple}
and \ref{twist}. In other words, one could attempt to implement
permutational quantum computation as a special case of anyonic
computation. As a more exotic possibility, it has been proposed that
in addition to Fermions and Bosons, which exchange according to
the two one-dimensional representations of the symmetric group, there
could also exist fundamental particles that exchange according to
higher dimensional representations of the symmetric group. Such
particles are said to obey parastatistics\cite{Peres}. Fundamental
particles obeying parastatistics have never been observed, but if they
were, they would provide a computational resource akin to
permutational quantum computing.

Finding more algorithms for permutational quantum computers provides
another direction for further research. In particular, it does not
seem obvious that the class of Ponzano-Regge transition amplitudes
approximated by the algorithm of section \ref{PR_algorithm} exhausts
the capabilities of permutational quantum computers. It would be
interesting to attempt to fully characterize the set of Ponzano-Regge
spin foams efficiently approximable in PQP.

\section{Acknowledgements}
In doing the work reported in this paper, I have benefitted from
conversations with numerous people. I especially thank Laurent
Freidel, Gorjan Alagic, and Liang Kong. I gratefully acknowledge
support from the Sherman Fairchild foundation and the National Science
Foundation under grant PHY-0803371, as well as the hospitality of the
Perimeter Institute.

\clearpage
\appendix

\section{Algorithm for two-manifold Equivalence}
\label{twoman}
Closed orientable two-manifolds are completely specified up to homeomorphism by
a single integer, the genus, which counts how many ``handles'' a
surface has, as illustrated below. \\
\begin{center}
\includegraphics[width=0.44\textwidth]{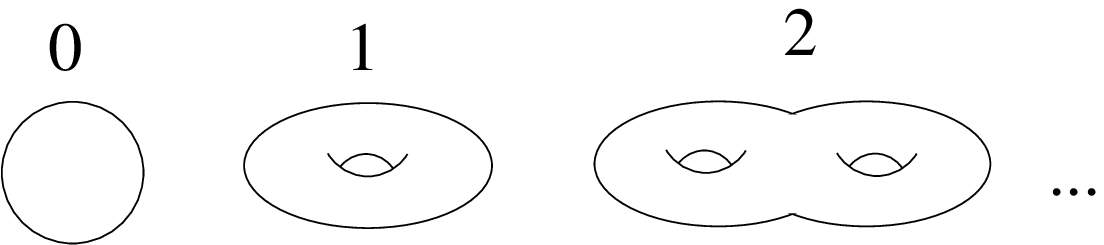} 
\end{center}
Genus is thus a complete two-manifold invariant; computing it solves the
2-manifold equivalence problem. Euler showed that any closed
triangulated orientable 2-manifold satisfies the formula 
\[
V-E+F=2-2g,
\]
where $V$ is the number of vertices, $E$ the number of edges, and $F$
the number of faces in the triangulation, and $g$ is the genus. Thus,
$g$ can be computed in time polynomial in the size of the
triangulation.

\section{Ponzano-Regge model as Quantum Gravity}
\label{gravity}
According to general relativity, spacetime is a manifold that locally
looks Minkowskian. That is, within a sufficiently small region of
spacetime there exists a choice of basis such that the distance metric
is
\[
\left[ \begin{array}{cccc}
-1 & 0 & 0 & 0 \\
 0 & 1 & 0 & 0 \\
 0 & 0 & 1 & 0 \\
 0 & 0 & 0 & 1 \end{array} \right]. 
\]
The vector component corresponding to the $-1$ matrix element is time,
and the three remaining components are space. Such a spacetime is
often referred to as $(3+1)$-dimensional to indicate the three spatial
dimensions and single time dimension.

One can formulate $(n+m)$-dimensional analogues of general relativity
for any integers $n$ and $m$. In particular, the $(3+0)$-dimensional
analogue of general relativity, which uses the ordinary Riemannian
metric
\[
\left[ \begin{array} {ccc}
1 & 0 & 0 \\
0 & 1 & 0 \\
0 & 0 & 1
\end{array} \right]
\]
turns out to be topological\cite{Baez}. Due to the difficulty of
formulating physically realistic quantum gravity, some physicists have
chosen to investigate quantum versions of $(3+0)$-dimensional
gravity. Due to its topological nature and other mathematical
conveniences, this seems to be easier than the full $(3+1)$-dimensional
case. Thus $(3+0)$-dimensional quantum gravity may serve as a useful
toy model on which to develop intuitions and techniques needed to
approach the full $(3+1)$-dimensional case.

The Ponzano-Regge model\cite{PR} is one proposed model of $(3+0)$-dimensional
quantum gravity. It is an example of a spin foam model. The spin foam
models are closely inspired by, but distinct from, loop quantum
gravity theories, which are obtained by applying canonical
quantization methods to general relativity. Interestingly, a
completely different motivation for the use of spin-networks in
describing gravity, not related to canonical quantization, was earlier
given by Penrose\cite{Penrose}.

\clearpage
\bibliography{pqp}

\end{document}